\documentstyle[11pt,aaspp4]{article}

\usepackage{epsf}

\newcommand{\rmn}{\rm}

\def\simless{\mathbin{\lower 3pt\hbox
   {$\rlap{\raise 5pt\hbox{$\char'074$}}\mathchar"7218$}}} 
\def\simgreat{\mathbin{\lower 3pt\hbox
   {$\rlap{\raise 5pt\hbox{$\char'076$}}\mathchar"7218$}}} 
\newcommand{\talbe}{T^{\alpha \beta}}
\newcommand{\galbe}{g^{\alpha \beta}}
\newcommand{\beq}{\begin{equation}}
\newcommand{\eeq}{\end{equation}}
\newcommand{\beqa}{\begin{eqnarray}}
\newcommand{\eeqa}{\end{eqnarray}}
\newcommand{\psupa}{p^{\alpha}}
\newcommand{\psupb}{p^{\beta}}
\newcommand{\usupa}{U^{\alpha}}
\newcommand{\usupb}{U^{\beta}}
\newcommand{\totpi}{\frac{2}{(2\pi)^3}}
\newcommand{\rr}{r^2}
\newcommand{\eo}{E_0}
\newcommand{\po}{P_0}
\newcommand{\fo}{F_0}
\newcommand{\rhoe}{\rho_e}
\newcommand{\pe}{P_e}
\newcommand{\ddr}{\frac{d}{dr}}
\newcommand{\sqggm}{\sqrt{\gamma^2-1}}
\newcommand{\ggm}{\gamma^2-1}
\newcommand{\pl}{\partial}

\newcommand{\gr}{\gamma(r)}
\newcommand{\tr}{T(r)}
\newcommand{\dlgdr}{\frac{d\ln\gamma}{dr}}
\newcommand{\dlgdlr}{\frac{d\ln\gamma}{d\ln r}}
\newcommand{\gi}{\gamma_i}
\newcommand{\siga}{\sigma_a}
\newcommand{\sigs}{\sigma_s}
\newcommand{\sigt}{\sigma_T}
\newcommand{\neeq}{n_{e,\rmn{eq}}}
\newcommand{\dne}{\delta n_e}

\newcommand{\req}{r_{\rmn{eq}}}
\newcommand{\rph}{r_{\rmn{ph}}}
\newcommand{\gameq}{\gamma_{\rmn{eq}}}

\newcommand{\teq}{T_{\rmn{eq}}}

\newcommand{\murf}{\mu_{\rmn{rf}}}

\newcommand{\fg}{f_{\gamma}}
\newcommand{\ual}{U^\alpha}
\newcommand{\ube}{U^{\beta}}
\newcommand{\tual}{TU_\alpha}
\newcommand{\tube}{TU_\beta}

\newcommand{\oalbe}{\omega_{\alpha \beta}}

\newcommand{\edot}{\dot{E}}
\newcommand{\mdot}{\dot{M}}
\newcommand{\ndot}{\dot{N}}

\def\nbbq{f_{\rmn{BB}}(q/\gamma T)}
\def\ngamqlamr{f_\gamma(q,\lambda,r)}
\def\ngamqlamrp{f_\gamma(q,\lambda^\prime,r)}
\def\lamp{\lambda^\prime}

\def\parr{{\partial\over\partial r}}
\def\parlam{{\partial\over\partial\lambda}}
\def\parq{{\partial\over\partial q}}

\def\ne{n_e}

\def\murf{\mu_{\rmn{rf}}}

\def\krf{k_{\rmn{rf}}}
\def\krfo{\krf^{(1)}}
\def\murfo{\murf^{(1)}}

\def\ktil{\tilde k}
\def\ktilo{\ktil^0}
\def\nhat{{\bf\hat n}}
\def\nhatprime{\nhat^\prime}
\def\fbb{f_{\rmn{BB}}}

\def\parr{{\partial\over\partial r}}
\def\parlam{{\partial\over\partial\lambda}}
\def\parq{{\partial\over\partial q}}

\def\ne{n_e}

\def\murf{\mu_{\rmn{rf}}}

\def\krf{k_{\rmn{rf}}}

\newcommand{\tl}{t_{\rmn s}}

\def\gbar{\overline{\gamma}}
\def\Hbar{\overline{H}}
\def\Tbar{\overline{T}}
\def\Vbar{\overline{V}}
\def\Ubar{\overline{U}}

\def\vbar{\overline{v}}
\def\ellp{\ell(\ell+1)}
\def\tl{\tau_\ell}
\def\pell{P_\ell}
\def\kl{k_\ell}
\def\gradvec{{\mbox{\boldmath $\nabla$}}}
\def\vvec{{\bf v}}
\def\Evec{{\bf E}}
\def\Bvec{{\bf B}}

\def\evec{{\bf\hat e}}
\def\er{\evec_r}
\def\eth{\evec_\theta}
\def\ephi{\evec_\phi}
\def\dpsi{{d\psi(\theta)\over d\theta}}
\def\Svec{{\bf S}}

\begin{document}

\title{Non-equilibrium effects in steady relativistic $e^+e^-\gamma$ winds}
\author{Ole M. Grimsrud and Ira Wasserman}
\affil{Center for Radiophysics and Space Research, Cornell University, Ithaca, New York 14853}

\begin{abstract}

We consider an ultra-relativistic wind consisting of 
electron-positron pairs and photons with the principal
goal of finding the asymptotic Lorentz factor $\gamma_{\infty}$ 
for zero baryon number. 
The wind is assumed to originate at radius $r_i$ where it has 
a Lorentz factor $\gamma_i$ and a temperature $T_i$ 
sufficiently high to maintain pair equilibrium.
As $r$ increases, $T$ decreases and becomes less than the 
temperature corresponding to the electron mass $m_e$, 
after which non-equilibrium effects become important.
Further out in the flow the optical depth $\tau$ drops below one,
but the pairs may still be accelerated by the photons until $\tau$ falls
below $\sim 2\times10^{-5}\gamma_{i}^{3/4}$. 
Radiative transfer calculations show that 
only at this point do the radiation flux
and pressure start to deviate significantly from their 
blackbody values.
The acceleration of the pairs increases $\gamma$ by a factor 
$\sim 45$ as compared to its value at the photosphere;
it is shown to approach $\gamma_{\infty} \sim
1.4\times 10^3 (r_i/10^6\mbox{cm})^{1/4} \gamma_{i}^{3/4} T_i/m_e$.

The limit of zero baryon number is a good approximation when the 
mass injection rate $\mdot$ in the flow is below a critical value
corresponding to 
$(\edot/\mdot)_{\rmn{c,0}}\sim 5\times10^7(r_i/10^6\mbox{cm})T_i/m_e$
for fixed energy injection rate $\edot$. For large baryon loading,
$\edot/\mdot\simless(\edot/\mdot)_{\rmn{c,M}}\sim
350(r_i/10^6\mbox{cm})^{1/4}\gamma_{i}^{3/4}T_i/m_e$, 
the asymptotic Lorentz factor is $\gamma_\infty\sim\edot/\mdot$.
Surprisingly, increasing $\edot/\mdot$ from $(\edot/\mdot)_{\rmn{c,M}}$ 
to $\infty$ only increases $\gamma_\infty$ by 
a factor $\sim (m_p/m_e)^{1/4}\approx 6.5$, 
less than an order 
of magnitude.

\end{abstract}


\section{Introduction}
\label{intro}

The release of a large amount of radiative energy into a 
small volume can lead to the formation of a fireball, a dense 
fluid of radiation and particles that expands under its own pressure.
Fireball models have become the accepted framework for understanding
gamma--ray bursts at cosmological distances and their afterglows
(Paczy\'{n}ski 1986, 1990; Shemi \& Piran 1990; 
M\'{e}sz\'{a}ros \& Rees 1993, Piran 1997). 
Paczy{\'n}ski (1986) and Goodman (1986) originally considered 
the possibility that fireballs could originate in the collision
of a pair of neutron stars in a binary star system coalescing 
as a result of gravitational radiation reaction. 
(See also Naryan, Paczy{\'n}ski \& Piran 1992.)
In this picture,
the thermal energy released in the collision, $\sim 10^{53}$ ergs,
is radiated as a neutrino--anti-neutrino burst. A fraction of 
that energy may be transformed into electron-positron pairs
above the surface of the neutron star (Goodman, Dar \& Nussinov 1987).
There are now additional proposals for the origin of fireballs
(e.g. Paczy{\'n}ski 1997; Fuller \& Shi 1997; Pen, Loeb \& Turok 1997).

Close to the radius at which energy is injected, the resulting 
wind is opaque.
The radiation
energy that is initially trapped can escape in two different ways
further out in the flow:
When the plasma becomes optically thin, radiation streams freely 
to the observer (Paczy{\'n}ski 1990). Alternatively,
if there is a significant baryon contamination in the fireball, 
it can become matter dominated before
radiation escapes. The matter will increase the opacity and, more importantly,
convert part of the radiation energy into bulk kinetic energy
(Shemi \& Piran 1990).
Interactions between the expanding atmosphere and the surrounding 
matter provide a way to convert the kinetic energy
in the baryons back to radiation at the resulting shock front
(M\'{e}sz\'{a}ros \& Rees 1993).
Internal shocks due to variations in the velocity of matter 
is also proposed as a way to dissipate kinetic energy
(Rees \& M\'{e}sz\'{a}ros 1994).
Fenimore (1997) recently pointed out that the observed
temporal structure of gamma ray bursts severely constrains the 
proposed models of energy conversion by relativistic shocks;
conceivably internal structure and shocks can account for some
of the observed variability (e.g. Kobayashi, Piran \& Sari 1997).

In the fireball models invoked to explain the afterglows of 
GRB 970228 and GRB 970508, the bulk Lorentz factor 
of the outflow is $\gamma \sim 100-1000$ before the outgoing 
shell is slowed significantly by sweeping up matter from ambient gas
(Wijers, Rees \& M\'{e}sz\'{a}ros 1997; Waxman 1997; Waxman, Kulkarni \&
Frail 1998).
From a theoretical point of view, values of $\gamma$ in this range 
yield acceptable estimates of burst duration and (with additional 
assumptions) characteristic photon energies.
It is presumed that most of the energy originally in the fireball 
converts to kinetic energy long before deceleration begins, so 
the (baryon) rest mass of the flow must be nonzero: 
$\mdot=\edot/\gamma$, where $\mdot$ and $\edot$ are the rest mass and 
total energy injection rate of the flow. Precisely how such small but nonzero
$\mdot/\edot$ arises is not clear yet; nor is it obvious whether 
$\gamma=\edot/\mdot$ can be much larger or smaller than 100-1000, the
values that seem necessary for modeling gamma ray bursts.

In this paper we reconsider the original steady wind problem 
first solved by Paczy{\'n}ski (1986) but for $\mdot\equiv 0$.
At first sight, one might think that the result would be 
$\gamma \to \infty$. 
However, the failure of equilibrium at low temperatures 
(once the pair density falls sufficiently so that annihilation 
becomes slow) leads ultimately to a finite $\gamma$. 
As we shall see,
in regions where the temperature is greater than the electron mass,
the fireball is very optically thick because of the large number of
electron--positron pairs. As the radius increases and the 
local temperature decreases,
the deviation from equilibrium in the number density
of pairs becomes significant.
A little further out in
the flow the optical depth falls below unity. 
However, the remaining pairs are still heated and accelerated 
considerably via their interactions with the radiation field; 
in the rest frame of the pairs,
the radiation field itself remains close to the blackbody form
long after the fireball becomes optically thin.
The radiation spectrum detected by a stationary observer would 
not be blackbody, however, but also differs from the power-law spectra
of {\sc grb}s.
Although our calculations pertain to a steady, spherical wind,
the general result that $\gamma$ is finite even at zero baryon 
mass may be true as well for thin shells emitted from impulsive 
energy release. We consider non-spherical perturbations around
our wind solutions in Section {\ref{pert}}; as we shall see,
some memory of surface `hot-spots' may persist out to the 
photosphere.

The dynamics for winds with sufficiently small baryon number
are similar to $\mdot=0$ outflow. 
A larger baryon number will increase the inertia of the flow,
and so we expect the radiative acceleration and therefore the
asymptotic Lorentz factor $\gamma_\infty$ to be reduced.
However, as the baryon number increases, the photospheric 
radius tends to increase as well, with a corresponding 
increase in the Lorentz factor at the photosphere.
These two factors combined result in a surprisingly small
variation in $\gamma_\infty$: For fixed $\edot$, as
$\edot/\mdot$ decreases from $\infty$ to 
$(\edot/\mdot)_{\rmn{c,M}}\sim
350(r_i/10^6\mbox{cm})^{1/4}\gamma_{i}^{3/4}T_i/m_e$,
$\gamma_\infty$ is only reduced by a factor of $\sim 10$.
For even smaller $\edot/\mdot$, the baryons will dominate the
energy of the flow even inside the photosphere, which 
results in a final Lorentz factor $\gamma_\infty\sim\edot/\mdot$.

We present a detailed analytical model of the dynamics
of the fireball in Section \ref{secan}. This includes approximate 
results for the 
asymptotic value of the Lorentz factor and of
the energy content in the pairs relative to that of the radiation,
based on the initial temperature and initial velocity of the flow. 
Furthermore, we estimate where the pairs go out of equilibrium,
the position of the photosphere, and the radius and optical depth 
at which the
radiation fields start to deviate from their blackbody values.
In Section \ref{results} we show results from a numerical 
calculation to which the analytical model is compared.
Next, based on the equation of radiative transfer, the 
comoving frame photon
distribution function is shown to be very close to blackbody 
even out to quite small optical depths in these ultra-relativistic flows.
The dynamical importance of baryons is described in Section \ref{baryon}.
There we obtain approximate results in different regimes characterized 
by the amount of baryon loading, and we integrate the dynamical
equations with baryons included in order to show how they affect 
the asymptotic Lorentz factor.
Finally, we discuss qualitatively several possible extensions
of our model, including the effects of (1) muon pairs and even 
nucleon pairs which could be present at sufficiently high 
temperatures; (2) a temperature anisotropy at the inner boundary,
and (3) magnetic fields.

\section{Equations}
\label{equations}

Consider a situation in which a large amount of energy is released 
into a compact region and the resulting relativistic outflow 
expands into vacuum. 
For simplicity we assume spherical symmetry and a stationary flow; 
gravity may also be neglected 
since we are interested in super-Eddington luminosities.
(Perturbations away from exact spherical symmetry will be considered 
in Section {\ref{pert}}.)
The cross-sections for absorption and scattering, $\siga$ and $\sigs$, are 
taken to be constants; letting $\siga$ and $\sigs$ cover a wide range
of values in the numerical calculations facilitates a qualitative
understanding of the effects of matter-radiation interactions.
Throughout, units for which $k_B=\hbar=c=1$ are used.

The temperature at the radius where energy is released is assumed to 
be high enough for pair creation to occur, and the flow is
very optically thick close to its inner boundary.
When the optical depth $\tau$ is $\gg 1$, the radiation field 
in the comoving frame will be close 
to blackbody. In a static or slowly expanding atmosphere one would expect 
the deviations from blackbody in the radiation field
to be of order $1/\tau$. However, in an ultra-relativistic (Lorentz factor 
$\gamma \gg 1$) expanding 
atmosphere, the corrections to blackbody actually vanish to first 
order in $1/\tau$ and lowest order in $1/\gamma$. 
This will be discussed further in Section \ref{discussion}.

In the opposite limit, the optical depth approaches zero and the radiation
streams freely. 
In this case a photon preserves its
frequency in the lab frame as well as the quantity $(1-\mu^2)r^2$, where
$\mu$ is the direction cosine relative to the local radial direction. 
As the radial coordinate $r$ gets large, the intensity therefore 
becomes increasingly sharply peaked in the outward radial direction;
this is a purely geometrical effect (Hummer \& Rybicki 1971).
In a static or slowly moving atmosphere one would expect the radiation 
quantities to develop large deviations from their optically 
thick values soon after reaching the regime where $\tau<1$. However,
we shall find that in a rapidly expanding atmosphere the radiation 
field maintains its equilibrium form long after it decouples from
the matter. This is similar to what happens in expanding universe
cosmology, where the background radiation preserves its blackbody 
spectrum after decoupling.
Specifically, as will be discussed in detail later 
(Section \ref{am_tault1}), we find that the
radiation fields start deviating from blackbody radiation 
in the comoving frame only
when $\tau \leq 2\times10^{-5}\gamma_{i}^{3/4}$.

This suggests a simple model for the radiation quantities:
For $\tau > 1$, the radiation field is approximated as
a blackbody distribution in the
rest frame of the flow; for $\tau < 1$, the radiation energy,
flux and pressure are given by the free-streaming approximation.
We also need equations for the matter, characterized by
its temperature $T$, Lorentz factor $\gamma$, and pair number density $n_e$.
Energy and momentum conservation give two dynamical
equations, whereas the Boltzmann equation determines the
number density for the electrons and positrons.

\subsection{Equations for $\tau>1$}
We assume that the flow originates at radius $r_i$ where it has a Lorentz 
factor $\gamma_i$ and temperature $T_i$.
The basic equations describing the flow are found from the energy-momentum tensor,
\beq
\talbe=\totpi\int\psupa\psupb f\frac{d^3p}{E}.
\label{eq1}
\eeq
$\psupa$ is the four-momentum, $f$ the distribution function, and $E=\sqrt{p^2+m^2}$ with $m$ being the mass of the particle contributing to $\talbe$.
For the matter ($e^\pm$ pairs for most of this paper)
\beq
\talbe_M=(\rho_M+P_M)\usupa \usupb + P_M g^{\alpha \beta},
\eeq
where $\rho_M$ and $P_M$ are the energy density and pressure 
measured in the rest frame of the flow and $\usupa$ is the matter 
four-velocity.
We also need the following expressions for number density $n$, 
energy density $\rho$ and pressure $P$:
\beqa
n & = & \totpi\int f(p,T)4\pi p^2 dp; \label{eq2} \\
\rho & = & \totpi\int \sqrt{p^2 + m^2} f(p,T)4\pi p^2 dp 
\;\;\;\;\;\;\;\; \mbox{and} \label{eq3}   \label{eq3.5} \\
P & = & \frac{1}{3} \totpi\int \frac{p^2}{\sqrt{p^2+m^2}} f(p,T)4\pi p^2 dp.
\label{eq4}
\eeqa

The non-zero components of the energy-momentum tensor for the radiation 
are $T_{\rmn co}^{00}=E_0, T_{\rmn co}^{01}=F_0, T_{\rmn co}^{11}=P_0$
and $T_{\rmn co}^{22}=T_{\rmn co}^{33}=(1/2)(E_0-P_0)$ in the comoving 
frame, where the 1-direction is along the flow velocity;
Lorentz transforming to an arbitrary reference frame gives 
\beqa
T_{R}^{00} & = & \gamma^2 [\eo + 2v\fo + v^2 \po]  \\
T_{R}^{01} & = & \gamma^2 [v(\eo + \po) + (1+v^2)\fo]  =  T_{R}^{10} \\
T_{R}^{11} & = & \gamma^2 [\po + 2v\fo + v^2 \eo]  \\
T_{R}^{22} & = & \frac{1}{2} (\eo - \po)  =  T_{R}^{33}.
\eeqa
Below, we call
$\eo,\po$ and $\fo$ the (comoving frame) radiation energy density, 
pressure and flux (e.g. Mihalas \& Mihalas 1984).

The equations describing the flow are given by
$\talbe_{;\beta }=0$,
where 
$\talbe \equiv \talbe_M + \talbe_R$.
With the assumptions of spherical symmetry and steady state, one gets
\beq 
\frac{1}{\rr } \ddr \left\{ 
\rr \left[ 
(\rho_M + P_M + \eo + \po ) \gamma \sqggm + ( 2 \gamma^{2} -1 ) \fo \right] 
\right\} = 0
\label{eq6} 
\eeq
and
\beq
\frac{1}{\rr }\ddr\bigl\{ \rr \bigl[ (\rho_M + P_M + \eo + \po )(\ggm) + 2\gamma\sqggm \fo \bigr] \bigl\} + \frac{dP_M }{dr} + \frac{d\po }{dr} + \frac{3\po - \eo }{r} = 0.
\label{eq7}
\eeq
For large $\tau$, the radiation has a blackbody distribution in the rest 
frame of the flow, and
$\eo=\pi^2T^4/15, \fo = 0$ and $\po=\eo/3$.

When the temperature in the rest frame of the fluid falls below $m_e$, the equilibrium $e^+ + e^- \rightleftharpoons 2\gamma$ can no longer be maintained. 
Consequently, the number densities of electrons and positrons will deviate 
from their equilibrium values and must be found from the Boltzmann equation,
\beq 
\frac{\psupa}{m_e}\frac{\pl f}{\pl x^{\alpha}} 
= \left( \frac{\pl f}{\pl t} \right)_{\rmn{collisions}}.
\label{eq8}
\eeq
Integrating over $d^3p$ 
we get, in spherical symmetry,
\beq
\frac{1}{\rr}\ddr \left\{ \rr n_e \gamma v \right\} = \left( \frac{\pl n_e}{\pl t} \right)_{\rmn{collisions}}.
\label{eq10}
\eeq
For the process 
\beq
e^- + e^+ \rightleftharpoons \gamma + \gamma^{\prime} ,
\label{eq11}
\eeq
the collision term is obtained by integrating the relevant matrix 
element times the factor $(\fg f_{\gamma^\prime} - f_-f_+)$ over 
phase space.
Deviations from pair equilibrium become significant when 
$T/m_e$ is small, so
we may approximate $f_i\ll1$ (i.e. $f_i = \exp\left[ -\left(E_i - \mu_i \right)/T
\right]$).
We can therefore neglect Fermi suppression 
of electron and positron final states and stimulated emission of
photons in the regions where equilibrium fails and the pair density 
`freezes out'.
As will be seen later (Section \ref{sec_aneq}), 
equilibrium is maintained until $T$ is $\sim 0.05m_e$.
With these approximations,
\beq
\left( \frac{\pl n_e}{\pl t} \right)_{\rmn{collisions}} =
 - \langle \sigma_{\rmn{ann}} v \rangle \left[ n_{e}^{2} - n_{e,\rmn{eq}}^{2} \right],
\label{eq14}
\eeq
where $n_{e,\rmn{eq}}$ is the equilibrium number density of electrons; 
using this result,
equation (\ref{eq10}) can be written
\beq
\frac{1}{r^2} \ddr \left\{ n_e r^2 \gamma v \right\} =
 - \langle \sigma_{\rmn{ann}} v \rangle \left[ n_{e}^{2} - n_{e,\rmn{eq}}^{2} \right].
\label{eq15}
\eeq
Svensson (1982) constructed a useful approximate expression for 
$\langle \sigma_{\rmn{ann}} v \rangle$, valid at all temperatures:
\beq
\langle \sigma_{\rmn{ann}} v \rangle  =  
\frac{ \pi r_{e}^{2} }{ 1+2(T/m_e)^2/
\ln\left[ 2\eta_ET/m_e + 1.3 \right] }
\eeq
Here $r_e$ is the classical electron radius, $\eta_E \equiv \exp(-C_E)$
and $C_E \approx 0.5772$ is Euler's constant.
For the purpose of our analytical estimates we only need to know that 
$\langle \sigma_{\rmn{ann}} v \rangle \approx \pi r_{e}^{2}$ for $T \ll m_e$.

\subsection{Equations for $\tau<1$}
\label{sub_tault1}

For small $\tau$, a good approximation to the radiation fields
is found by assuming that photons stream freely from the photosphere
to the observer. However, as the optical depth drops below one, 
energy and momentum may still be deposited by photons in the pair wind. Even though the energy and momentum
deposition may be small compared to the energy-momentum content of the escaping photons, it may be large compared to that of the pairs.
The dynamical equations must therefore take this interaction
into account. We shall see that substantial acceleration of the 
$e^\pm$ pairs occurs at $\tau<1$.

The equation of radiative transfer can be written schematically as
\beq
k^{\beta} \frac{\pl}{\pl x^{\beta}}\fg(k,x) = C_{\gamma}.
\label{eq16}
\eeq
$\fg(k,x)$ is the phase space distribution function of the 
photons, and the Lorentz invariant 
collision terms $C_\gamma$ result from interactions with the electrons and 
positrons in the fluid.

We are interested in three processes: $e^{\pm}-\gamma$ scattering
($e^{\pm} + \gamma \to e^{\pm} + \gamma$),
bremsstrahlung and its inverse
($e^{\pm} + e^{\pm} \rightleftharpoons e^{\pm} + e^{\pm} + \gamma$),
and pair annihilation and creation
($e^+ + e^- \rightleftharpoons 2\gamma$).
We use a phenomenological model for the collision terms, evaluated in
the rest frame of the flow. For scattering the collision term is
\beq
{\cal C}_s=2n_e\sigma_s \tilde k^0 \left[-\fg(k,x)
+\int{d^2\hat n^\prime\,g(\hat n^\prime\to\hat n)\fg(k^\prime,x)}\right]
\eeq
where $\tilde k^0$ is the photon energy in the rest frame of the flow, and
we assumed that the scattering is elastic ($\tilde k^{\prime\,0}=\tilde k^0$).
The probability distribution for direction changes in the rest frame is
$g(\hat n^\prime\to\hat n)$.
The cross section for scattering,  $\sigma_s$, is taken to be 
a constant.
The collision term for absorption and emission is 
\beq
{\cal C}_a=2n_e\sigma_a \tilde k^0[-\fg(k,x)+\fbb(\tilde k^0/T)],
\eeq
where $\fbb(\tilde k^0/T)$ is the equilibrium photon distribution function at 
temperature $T$. The absorption cross
section $\siga$ is also assumed to be constant.

Now take the first moment of the Boltzmann equation (\ref{eq16}):
\beq
\totpi \int \frac{d^3k}{k^0} k^{\alpha} \left[ k^{\beta} \frac{\pl}{\pl x^{\beta}}\fg(k,x) \right] = 
\totpi \int \frac{d^3k}{k^0} k^{\alpha} C_\gamma.
\label{eq17}
\eeq
We recognize the left hand side as $\talbe_{R;\beta}$. The right 
hand side is the radiation force, which we denote $G^{\alpha}$.
The physical interpretation of the equations 
$\talbe_{R;\beta}=G^{\alpha}$ (or $\talbe_{M;\beta}=-G^{\alpha}$)
is that $-G^0$ equals the net rate of radiative energy density input 
into the matter, while $-G^i$ is the net rate of radiative momentum input  
(Mihalas \& Mihalas 1984).
For the scattering term, taking the first moment over ${\cal C}_s$ as
indicated in equation (\ref{eq17}) gives
\beq
G_{\rmn{s,0}}^0=G_{\rmn{s,0}}^2=G_{\rmn{s,0}}^3=0;\qquad\qquad 
G_{\rmn{s,0}}^1=-2n_e\sigma_sF_0,
\eeq
where lower index 0 denotes that these quantities are evaluated in the
comoving frame.
The components of the radiation four-force corresponding to
absorption and emission are
\beq
G_{\rmn{a,0}}^0=-2n_e\sigma_a[E_0-U_{\rmn{eq}}(T)]\qquad\qquad G_{\rmn{a,0}}^1=-2n_e\sigma_a F_0
\qquad\qquad G_{\rmn{a,0}}^2=G_{\rmn{a,0}}^3=0,
\eeq
where $U_{\rmn{eq}}(T)=\pi^2T^4/15$. 
Finally, for pair annihilation and creation we use the interpretations 
of $G^{\alpha}$ to write down directly
\beq
G_{\rmn{p,0}}^0=2m_e\langle \sigma_{\rmn{ann}} v \gamma \rangle[n_e^2-n_{e,\rmn{eq}}^2(T)] \qquad\qquad
G_{\rmn{p,0}}^i=0\qquad\qquad[i=1,2,3].
\eeq
The energy loss of the plasma per annihilation is $(\gamma_- + \gamma_+)m_e$.
Svensson (1982) gives an approximate expression for the 
pair annihilation cooling rate:
\beq
\langle \sigma_{\rmn{ann}}v\gamma \rangle  =  \frac{ \pi r_{e}^{2} }
{ 1/(1+6T/m_e) + (T/m_e)/\left\{ \ln\left[ 2\eta_ET/m_e + 1\right] + 1/4 \right\} }.
\eeq
In the non-relativistic limit the energy loss is dominated by the rest mass
energy, and one gets $\langle \sigma_{\rmn{ann}}v\gamma \rangle \approx \pi r_{e}^{2}$ for $T\ll m_e$. 

Using the energy momentum tensor $\talbe_{M}$ for the pairs and 
the components of the four force $G^{\alpha}$ as given above, the dynamical
equations $\talbe_{M;\beta}=-G^{\alpha}$ reduce to
\beq
(\rho_M + P_M) \dlgdlr + \frac{dP_M}{d\ln r} = -\frac{r}{\gamma} G_{0}^{r}
\label{eeq21_5}
\eeq
and
\beq
\frac{d\rho_M}{d\ln r} + 2(\rho_M + P_M) + (\rho_M + P_M)
\left( 1+\frac{1}{\gamma^2-1} \right)
\dlgdlr = -G_{0}^{0}\frac{r}{\sqrt{\gamma^2-1}}.
\label{eeq21_5.1}
\eeq
Up to this point, no restriction to small optical depths has been 
invoked, and the dynamical equations can be applied at any $\tau$,
provided $f_\gamma(k,x)$ is determined.
For $\tau < 1$, we assume that radiation streams outward freely.
This is only marginally accurate at $\tau\sim 1$, but becomes 
progressively more precise with decreasing $\tau$.
However, since the photon intensity is sharply forward-peaked
in the lab frame, this approximation is better than one might
have expected even at $\tau \simless 1$. 
A non-interacting photon preserves its frequency in
the lab frame as well as the quantity $(1-\mu^2)r^2$,  where
$\mu$ is the direction cosine relative to the local radial
direction. The phase space density $f_\gamma(k,x)$ is conserved along any
photon ray path. Consequently, we need only to relate
the photon frequency and direction cosine in the comoving
frame at different points in the flow to find $\eo$, $\po$, and $\fo$
at small $\tau$.
In the extreme relativistic limit, photons 
are highly beamed in the forward direction as seen in the
lab frame. It is therefore convenient to define the quantity
\beq
\eta=2\gamma^2(1-\mu),
\eeq
in terms of which the direction cosine in the rest frame is
approximately
$\murf\approx(1-\eta)/(1+\eta).$
As $\eta$ spans the range $[0,\infty]$, $\murf$ spans the
range $[-1,+1]$.
From $2(1-\mu)r^2 = \eta r^2/\gamma^2 \approx$ constant, we get
\beq
\murf={1+\murfo-(1-\murfo)\zeta (r)\over
1+\murfo+(1-\murfo)\zeta (r)}
\eeq
where
\beq
\zeta (r) \equiv \left({r_1\gamma\over r\gamma_1}\right)^2.
\eeq
The label `1' denotes a reference point
in the flow,
taken to be $\tau\sim 1$ below,
where we shall assume that the distribution 
function is known and beyond which photons stream freely.
Note that if $\zeta (r)=1$ then $\murf=\murfo$, which means
that the direction cosine relative to the local radial direction does
not change as long as $\gamma\propto r$; however, as
$\zeta (r)\to 0$, the distribution becomes increasingly peaked
around $\murf=1$.
We shall see that $\gamma \propto r$ is maintained to quite 
small values of $\tau$.

In a similar way, the photon frequency $\krf$ in the rest
frame of the flow is
\beq
\krf={k(1+\eta)\over 2\gamma}=\krfo{\gamma_1\over\gamma}
\biggl({1+\eta\over 1+\eta_1}\biggr),
\eeq
or, introducing $\murf$ rather than $\eta$,
\beq
\krf=\krfo{\gamma_1\over\gamma}
\biggl[{1+\murfo+(1-\murfo)\zeta (r)\over 2}\biggr].
\eeq
Now assume that the photon distribution function
is $f_\gamma(\krfo,\murfo)$ at $r_1$.
If the photons flow freely outside $r_1$, then at any larger
value of $r$, the Liouville theorem implies that the phase
space distribution is still $f_\gamma(\krfo,\murfo)$. 
Only the mapping from $r_1$ to $r$ is needed to find $E_0, P_0$ and $F_0$.

Suppose the distribution function for the radiation at $r_1$ can be 
expanded in terms of orthonormal Legendre polynomials:
\beq
f_\gamma(\krfo,\murfo)=f_0(\krfo) + \murfo f_1(\krfo) + 
\left( \frac{3(\murfo)^2-1}{2} \right) f_2(\krfo)
\eeq
Defining
\beq
I_1\equiv\int{dq\,q^3 f_0(q)} \qquad
2J_1\equiv\int{dq\,q^3 f_1(q)} \quad \mbox{and} \quad
10K_1\equiv\int{dq\,q^3 f_2(q)},
\eeq
the radiation energy, flux, and pressure can be expressed as:
\beqa
E_0 & = & {4\pi\over 3}\left({r_1\gamma_1\over r\gamma}\right)^2
\left\{I_1[1+\zeta (r)+\zeta^2(r)]+J_1[1-\zeta^2(r)] + 
K_1[1-\zeta]^2 \right\} \label{neo_eq}  \\
F_0 & = & {4\pi\over 3}\left({r_1\gamma_1\over r\gamma}\right)^2
\left\{I_1[1-\zeta^2(r)]+J_1[1+\zeta^2(r)] +
K_1[1-\zeta^2] \right\} \label{nfo_eq}  \\
P_0 & = & {4\pi\over 3}\left({r_1\gamma_1\over r\gamma}\right)^2
\left\{I_1[1-\zeta(r)+\zeta^2(r)]+J_1[1-\zeta^2(r)]
+K_1[1+\zeta]^2\right\}. \label{npo_eq}  
\eeqa
We can check that in the stationary frame
$r^2F(r)={\rm
constant}$, consistent with energy conservation.

For our purposes, the reference radius $r_1$ will be taken to be
the position of the photosphere. The radiation
field is then given by an isotropic blackbody distribution
for $r\leq r_1$. 
(In Section \ref{spec}, we estimate deviations from a blackbody.)
We therefore
specialize to $f_1=f_2=0$ (i.e. $J_1=K_1=0$) in the following discussion.

As long as $\zeta (r)=1$, $E_0=3P_0$ and $F_0=0$. However, as $\zeta
(r)\to 0$, $P_0\to E_0$ and $F_0\to E_0$; only when $\gamma\to{\rm
constant}$ do $P_0$, $E_0$ and $F_0$ decrease like $r^{-2}$. Note that
one can rewrite the prefactor in equations (\ref{neo_eq}--\ref{npo_eq}) as
\beq 
\left({r_1\gamma_1\over r\gamma}\right)^2
={(r_1/r)^4\over\zeta(r)};
\eeq
consequently, when $\zeta (r)=1$, the rest frame energy density and 
pressure both decrease like $r^{-4}$. This is so even though there is
no interaction with matter. The situation is the same as in expanding
universe cosmology, where the radiation field maintains its equilibrium
form after it decouples from the matter entirely.
We see from the above that when the optical depth is small, the
critical function that determines the approach to $F_0/E_0=1$
and $P_0/E_0=1$ is $\zeta (r)$.

For the pairs, we always solve the annihilation equation
derived in the previous subsection.

\section{Analytical Model}
\label{secan}

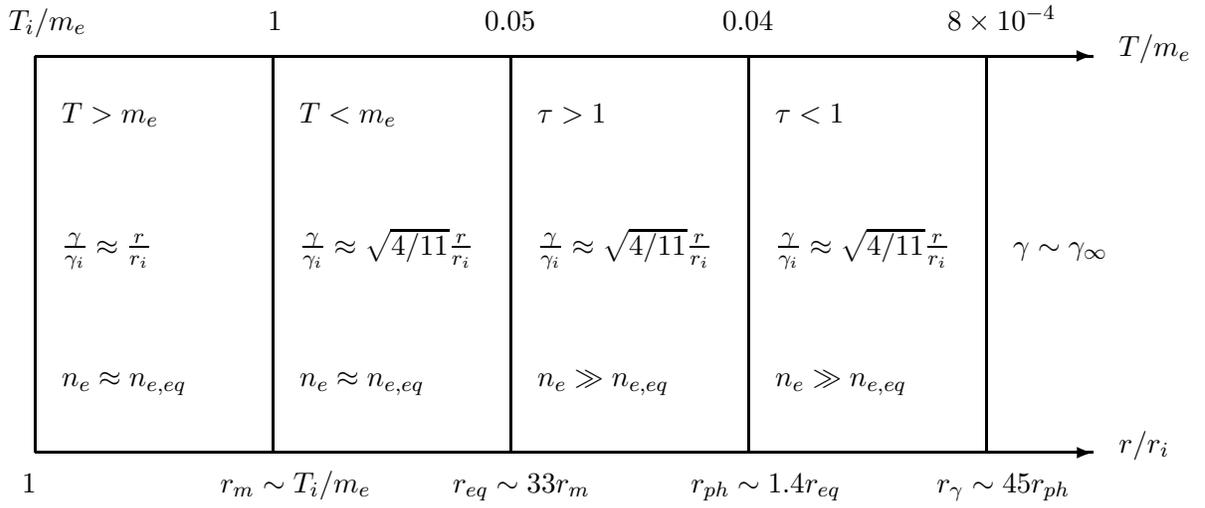
\begin{figure}

\begin{picture}(400,225)(-25,-25)
\thicklines
%
%
\put(0,0){\vector(1,0){400}}
\put(0,150){\vector(1,0){400}}
\put(0,0){\line(0,1){150}}
\put(90,0){\line(0,1){150}}
\put(180,0){\line(0,1){150}}
\put(270,0){\line(0,1){150}}
\put(360,0){\line(0,1){150}}
%
%
\put(-5,-15){1}
\put(70,-15){$r_{m}\sim T_i/m_e$}
\put(158,-15){$r_{eq}\sim 33 r_m$}
\put(248,-15){$r_{ph}\sim 1.4 r_{eq}$}
\put(341,-15){$r_{\gamma}\sim 45 r_{ph}$}
\put(410,0){$r/r_i$}
%
%
\put(-10,160){$T_i/m_e$}
\put(88,160){1}
\put(170,160){0.05}
\put(260,160){0.04}
\put(345,160){$8\times10^{-4}$}
\put(410,150){$T/m_e$}
%
%
\put(10,125){$T>m_e$}
\put(10,25){$n_e\approx n_{e,eq}$}
\put(10,75){$\frac{\gamma}{\gamma_i}\approx\frac{r}{r_i}$}
%
%
\put(100,125){$T<m_e$}
\put(100,25){$n_e\approx n_{e,eq}$}
\put(100,75){$\frac{\gamma}{\gamma_i}\approx\sqrt{4/11}\frac{r}{r_i}$}
%
%
\put(190,125){$\tau>1$}
\put(190,25){$n_e\gg n_{e,eq}$}
\put(190,75){$\frac{\gamma}{\gamma_i}\approx\sqrt{4/11}\frac{r}{r_i}$}
%
%
\put(280,125){$\tau<1$}
\put(280,25){$n_e\gg n_{e,eq}$}
\put(280,75){$\frac{\gamma}{\gamma_i}\approx\sqrt{4/11}\frac{r}{r_i}$}
%
%
\put(370,75){$\gamma\sim\gamma_{\infty}$}

\end{picture}
\caption{Summary of analytical model with $\gamma_i$ 
and $r_i/10^6\mbox{cm}$ both of order unity.
\label{summfig}}
\end{figure}

 In this section we will do back-of-the-envelope calculations and
obtain approximate results and scaling laws in four different 
regimes, each characterized by their local temperature $T$ and 
optical depth $\tau$: First, in the high temperature limit 
($T\gg m_e$), electrons, positrons and photons are in equilibrium, 
the optical depth is very large and we get
simple power laws for $\gr$ and $\tr$.
As the temperature drops below $m_e$, the number density of
electrons and positrons starts to deviate from its equilibrium value
and will be calculated approximatively from the annihilation 
equation.
In the region where the optical depth $\tau<1$, we
estimate how the radiation fields 
affect the velocity and the temperature of the matter.
Finally, we calculate the asymptotic values for $\gamma$, $Tr$ and
$n_e r^2$ for $\tau \to 0$.
A summary of the scaling laws can be found in Fig. {\ref{summfig}}.
The validity of the simple analytical model presented
here will be checked against a numerical solution of the 
equations in the following section.
There we will explore models with a range of different initial temperatures and
Lorentz factors.

\subsection{$\tau\gg 1$ and $T\gg m_e$}

When $T\gg m_e$, the radiation energy $\eo$
is given by the equilibrium Planck energy density, $\eo=(\pi^2/15)
T^4 \equiv aT^4 = 3\po$. 
The radiation flux is negligible since the optical depth 
$\tau\gg1$; we will 
check the consistency of this assumption later. (See Section 
\ref{tausub} for calculations of the optical depth.) \\
Introduce the notation
\beq
\rho \equiv \rho_M + \eo \equiv \rho_{e^-} + \rho_{e^+} + \eo 
\quad \hbox{and} \quad
P \equiv P_M + \po \equiv P_{e^-} + P_{e^+} + \po.
\eeq
Energy conservation, equation (\ref{eq6}), now simplifies to 
\beq
\ddr \left\{ r^2 (\rho+P)\gamma\sqrt{\ggm} \right\} = 0, 
\quad \hbox{i.e.} \quad
L \equiv 4\pi r^2\gamma\sqrt{\ggm}(\rho+P) = \hbox{constant}.
\label{eqa1}
\eeq
The Euler equation, equation (\ref{eq7}), becomes
\beq
(\rho+P)\dlgdr + \frac{dP}{dr} = 0,
\label{eqa2}
\eeq
where we made use of energy conservation (equation [\ref{eqa1}]).

When $\rho$ and $P$ depend on $T$ alone, 
$dP/dT=(\rho+P)/T$ (e.g. Weinberg 1972, Section 15.6), 
and the flow equations give
\beq
\frac{d\ln(\gamma T)}{d\ln r} = 0; \quad \gamma T = \hbox{constant} 
 = \gamma_i T_i,
\label{eqa3}
\eeq
and
\beq
\dlgdlr = \frac{2}{(Td\rho/dT)/(\rho+P) - 
\gamma^2/(\gamma^2-1)} =
\frac{2(\gamma^2-1)}{2\gamma^2-3}.
\label{eqa4}
\eeq
The last step in the above equation is valid for $\rho=3P \propto T^4$.
For $\gamma\gg1$, $d\ln\gamma/d\ln r=1$, i.e. 
\beq
\gamma \propto r; \quad T\propto \frac{1}{r}; \quad \gamma T = \gamma_i T_i
\quad \hbox{for $T\gg m_e$}
\label{eqa5}
\eeq
(e.g. Goodman 1986; Paczy\'{n}ski 1986).
Note that equation (\ref{eqa4}) requires a minimum initial velocity for the 
flow to lift off: 
Demanding $d\ln\gamma/d\ln r > 0$ implies that $\gi > \sqrt{3/2}$.

\subsection{$\tau\gg 1$ and $T\sim m_e$}
\label{sec_aneq}

When $T\gg m_e$, $\rho_M (=3P_M)$ is comparable to $\eo (=3\po)$ and
also proportional to $T^4$, but 
for $T\ll m_e$, $\rho_M\ll\eo$. In this subsection we want to explore 
what happens when $T\simless m_e$.
Define
\beq
h(T)\equiv \rho_M + P_M + \eo + \po \approx 
\left\{ \begin{array}{ll}
        \frac{11}{3}\frac{\pi^2}{15}T^4 & \mbox{for $T\gg m_e$} \\
        \frac{4}{3}\frac{\pi^2}{15}T^4 & \mbox{for $T\ll m_e$}.
        \end{array}
\right.
\label{am_hoft}
\eeq
In general, $\rho_M$ and $P_M$ should be found from equations 
(\ref{eq3.5}) and (\ref{eq4}).

We can now calculate the radius $r_m$ where the temperature equals
the electron mass, using 
conservation of energy:
\beq
\frac{r_m}{r_i} = \sqrt{ \frac{\gi \sqrt{\gi^2-1}}{\gamma_m \sqrt{\gamma_{m}^{2}-1}} \frac{h(T_i)}{h(m_e)} } 
 = \left( 1-\frac{1}{\gi^2} \right)^{1/4}\frac{m_e}{T_i} 
\sqrt{ \frac{h(T_i)}{h(m_e)} } \approx 
\left( 1-\frac{1}{\gi^2} \right)^{1/4} \frac{T_i}{m_e}
\label{am_rmi}
\eeq
Here $\gamma_m \gg 1$ and $\gamma_mm_e \approx \gi T_i$ 
was used.
Note that the pair density is still close to equilibrium when
$T\sim m_e$, which will be verified later in this subsection.

As $T$ drops below $m_e$, the equilibrium density of electrons and
positrons falls off exponentially as
\beq
\neeq \approx \frac{2}{(2\pi)^{3/2}} (m_eT)^{3/2} \exp(-m_e/T).
\eeq
Consequently, for $T\ll m_e$, $\rhoe, \pe \ll \eo, \po$, and
$\rho \approx \eo \approx 3P \approx 3\po \approx (\pi^2/15)T^4$.
But then we get the same scaling laws as in the high
temperature limit (see the justification of equations 
[\ref{eqa3}]-[\ref{eqa5}]):
\beq
\gamma \propto r; \quad T\propto \frac{1}{r}; \quad \gamma T = \mbox{constant}
\quad \hbox{for $T\ll m_e$}.
\eeq
As long as equilibrium is maintained,
the equation of state of the flow is purely a function of temperature,
and the Euler equation implies that $\gamma T$ = constant.
At the same time, energy conservation requires
\beq
L \approx 4\pi r^2h(T) \gamma \sqrt{\gamma^2-1} \approx 
4\pi r^2 (\gamma_i T_i)^2 T^2 \alpha  =  \mbox{constant}.
\eeq
Thus, $r^2 T^2 \alpha$ = constant, resulting in the scaling laws
\beqa
&\gamma \approx \gamma_i(r/r_i)  \qquad \mbox{and}   \qquad 
 T \approx T_i(r_i/r) &  \qquad \mbox{for $T\gg m_e$};   \label{am_gshift1} \\
&\gamma \approx \gamma_i(r/r_i)\sqrt{4/11}  \qquad \mbox{and}   \qquad 
 T\approx T_i(r_i/r)\sqrt{11/4}  & \qquad \mbox{for $T\ll m_e$}. 
\label{am_gshift2}
\eeqa
The change in slope is due to the fact that pairs contribute 
little to $\rho + P$ once they become non-relativistic.
This remains true for small $T<m_e$ even though $e^\pm$ are 
far more plentiful than in pair equilibrium.

At somewhat larger radius than $r_m$, deviations from pair annihilation
equilibrium will become important. 
(Lee \& Weinberg (1977) considered the analogous problem for heavy leptons
in the early universe.)
In order to estimate where this happens,
let us assume that the deviations are small, i.e. $\dne \equiv
n_e - \neeq \ll \neeq$. Then expand
$n_{e}^{2}-\neeq^2 \approx 2\neeq \dne$
in the pair equation (\ref{eq15}) to get
\beq
\dne \approx -\frac{1}{2 r^2 \langle \sigma_{\rmn{ann}} v \rangle \neeq}
\ddr \left\{ \neeq r^2 \sqrt{\ggm} \right\}.
\eeq
Since $\gamma \propto r$ as long as equilibrium is maintained, we get
\beq
\dne \approx -\frac{\gamma}{2 r \langle \sigma_{\rmn{ann}} v \rangle}
\left[ 3 + \frac{d\ln \neeq}{d\ln r} \right].
\eeq
From the exponential form of $\neeq$,
$d\ln \neeq/d\ln r = - d\ln \neeq/d\ln T
\approx -m_e/T$
for $T\ll m_e$.

Equilibrium fails when $\dne$ becomes comparable to $\neeq$.
Let $\req$ (and $\teq$) be the radius (and temperature) 
where $\dne = \neeq$; then
\beq
\frac{2}{(2\pi)^{3/2}} (m_e\teq)^{3/2} \exp(-m_e/\teq) =
\frac{\gameq}{2\req\pi r_{e}^{2}} 
\frac{m_e}{\teq}
\eeq
to leading order in $m_e/\teq$.
Relate $\teq$ and $\req$ by $\teq \req \sim \sqrt{11/4}m_e r_m$, 
and use $\gameq/\req \approx \sqrt{4/11} \gi/r_i$ (from the limiting 
cases of $h(T)$, assuming $\teq\ll m_e$ and $T_i > m_e$, and 
$\gamma\sqrt{\gamma^2-1}\approx \gamma^2$ in the energy conservation equation).
We then find
\beq
\frac{2}{(2\pi)^{3/2}} m_{e}^{3} \left( \frac{11}{4} \right)^{3/4}  
\left( \frac{r_m}{\req} \right)^{3/2} 
\exp(-\sqrt{4/11}\req/r_m) =
\frac{ (4/11)(\gi/r_i) }{2\pi r_{e}^{2}} 
\frac{\req}{r_m},
\eeq
which has the approximate solution
\beq
\req/r_m \approx \sqrt{11/4}\ln \left[ 
\sqrt{2/\pi} (11/4)^{7/4} (r_{e}^{2} r_i m_{e}^{3})/\gi
\right] -
\sqrt{11/4}(5/2)\ln \left[ \frac{\req}{r_m} \right].
\label{am_reqm}
\eeq
Note that $\req/r_m$ does not depend on the initial temperature $T_i$ 
and is only weakly dependent on $\gamma_i$.
For $r_i=10^6$cm, the numerical value is
$\req/r_m \approx 33-\sqrt{11/4}\ln\gamma_i$.
The temperature at $\req$ is approximately $\teq \sim \sqrt{11/4}m_e r_m/\req 
\sim 0.052m_e$.

\subsection{$\tau\sim 1$ and $T\ll m_e$}
\label{tausub}

At $r_{\rmn{eq}}$ the energy density of the flow is dominated by the photons:
Since $\teq \ll m_e$ we have
\beq
\frac{\rho_{\rmn{M,eq}}}{E_{\rmn{0,eq}}} \approx \frac{2m_e n_e(\req)}{(\pi^2/15)\teq^4}
 \approx 1.1\times 10^{-6}\gamma_i.
\eeq
As long as the flow remains photon dominated and the radiation field
stays close to its blackbody form, $\gamma \propto r$.

Now, let us estimate the optical depth in the flow. 
In the rest frame of the flow, the density of electrons
and positrons is $2n_e \equiv n_{-} + n_{+}$ and the cross 
section for interaction is $(\sigs + \siga)$.
A reasonable choice for the optical depth of the
flow is the radial depth (Abramowicz, Novikov \& Paczy\'{n}ski 1991)
\beq
\frac{d\tau}{dr} = 2n_e(\sigs + \siga) \sqrt{1-v\over 1+v} \approx 
\frac{n_e (\sigs + \siga)}{\gamma},
\eeq
where the last approximation holds for an extremely relativistic flow.
We then find
\beq
\tau(r) = \frac{8\pi}{3} r_{e}^{2} \left(\frac{\siga}{\sigt} + \frac{\sigs}{\sigt}\right)
 \int_{r}^{\infty}dr\frac{n_e(r)}{\gamma(r)}.
\eeq
As long as 
the radiation field remains close to a blackbody, $\gamma(r)\approx\gameq(r/\req).$ 
Since $n_e > \neeq$ outside $\req$, we can approximate the pair
equation as
\beq
\frac{1}{r^2} \ddr \left( n_e r^3 \right) \approx
 -\pi r_{e}^{2} \frac{\req}{\gameq} n_{e}^{2}.
\eeq
In other words, although pair annihilation continues outside $\req$, pair creation becomes unimportant as the number
of photons energetic enough to produce $e^\pm$ pairs decreases exponentially.
The solution to this equation is
\beq
n_e(r) \approx \frac{n_{e}(r_{\rmn{eq}})}{1+(1/3)\sqrt{4/11}
(\req/r_m) \left[ 1-
\left( \req/r \right)^3 \right]} \left( \frac{\req}{r} \right)^3
\label{am_n}
\eeq
for $r > \req$ and $\gamma \propto r$.
Since $n_e(\req) 
\sim \neeq (\req) + \delta n_e(\req) 
\sim 2\delta n_e(\req) 
\sim (4/11)(\gamma_i/r_i)(\req/r_m)/\pi r_{e}^{2}$,
\beqa
\tau(r) &\approx & \frac{8\pi}{3} r_{e}^{2} \left(\frac{\siga}{\sigt} + \frac{\sigs}{\sigt}\right)
\frac{\req}{\gameq} n_{e}(r_{\rmn{eq}}) \req^3 
\int_{r}^{\infty} \frac{dr}{ r^4 
\left\{ 
	1 + (1/3)\sqrt{4/11}(\req/r_m) \left[
        1-\left( \req/r \right)^3 \right]
\right\}  }  \nonumber \\
 & = & -\frac{8}{3} \left(\frac{\siga}{\sigt} + \frac{\sigs}{\sigt}\right) \ln\left\{ 
1- \frac{ (1/3) \sqrt{4/11} 
(\req/r_m) \left( \req/r \right)^3 }
{ 1 + (1/3) \sqrt{4/11} (\req/r_m) } \right\} 
\eeqa
at $r>r_{\rmn{eq}}$.
The photospheric radius $r_{\rmn{ph}}$ may be estimated from 
$\tau(r=r_{\rmn{ph}}) = 1$, implying
\beq
\frac{\rph}{\req} = \left\{ \left( 3\sqrt{\frac{11}{4}}\frac{r_m}{\req} + 1 \right) 
\left[ 1-\exp\left( -\frac{3\sigt}{8[\siga+\sigs]}\right) \right] 
\right\}^{-1/3}
\label{am_rpheq}
\eeq
For example, if $\siga+\sigs \sim \sigt$ then $\rph/\req \sim 1.4$.
This implies that
the temperature at the photosphere is $T_{\rmn ph}/m_e \approx 
(\req/\rph) T_{\rmn eq}/m_e \sim 0.037.$ 
The optical depth at $r_{\rmn{eq}}$ is $\tau_{\rmn{eq}} \sim 5.4 ( \siga + \sigs)/\sigt $.

Outside $\rph$ the photons stream nearly freely out to the observer.
The escaping photons form a high energy gamma ray continuum since
$T\gamma$, the temperature seen by a distant observer in the lab frame,
is approximately constant in the flow out to $\rph$.

\subsection{$\tau < 1$ and $T \ll m_e$}
\label{am_tault1}

For $\tau<1$, 
energy and momentum may still be deposited by photons in the pair wind.
Thus, the pairs can be
accelerated and heated via their interactions with escaping photons even though the wind is not opaque. Because the energy and momentum in pairs is 
smaller than in radiation at $\tau\sim 1$, even small depositions of 
energy and momentum may affect the pairs significantly.
In this regime, each electron or positron experiences a force
\beq
f_{\rmn{rf}}^r \approx (\siga + \sigs)\fo
\eeq
in the flow rest frame. 
(Since $T\ll m_e$, there is little difference between rest frames
of flow and individual electrons.)
Transforming to the
observer's frame implies 
\beq
{d\gamma\over dr} \approx (\siga + \sigs)\fo/m_e,
\eeq
where we used $d\tau=dt/\gamma$ and $dt=vdr\approx dr$.
This can also be seen from the dynamical equation (\ref{eeq21_5}).
The appropriate flux in this regime is given by
the free-streaming limit,
\beq
F_0  =  {4\pi\over 3}\left({r_1\gamma_1\over r\gamma}\right)^2
I_1[1-\zeta^2(r)],
\eeq
assuming, for simplicity, that the radiation field at $r_1$ is isotropic,
as will be the case when matching onto blackbody radiation at $\tau\sim 1$.
Define $z\equiv r/r_i$ and
\beq
\Lambda \equiv \frac{4\pi}{3}\left( \frac{\siga}{\sigt} + \frac{\sigs}{\sigt}
\right) (\sigt r_i m_{e}^{3}) \frac{I_1}{m_{e}^{4}} \frac{z_1}{\gamma_1} 
\approx 3.1\times 10^7 \left( \frac{\siga}{\sigt} + \frac{\sigs}{\sigt} \right)
\frac{1}{\gamma_i} \frac{r_i}{10^6\mbox{cm}} 
\left( \frac{\req}{\rph} \right)^4.
\eeq
Since $\gamma/\gamma_1=(z/z_1)\sqrt{\zeta}$, the 
equation of motion becomes
\beq
1+\frac{1}{2\zeta}\frac{d\zeta}{d\ln z} = 
\Lambda \left( \frac{z_1}{z} \right)^4 \zeta^{-3/2} (1+\zeta)(1-\zeta).
\label{am1}
\eeq
At the photosphere, $z=z_1$ and $\zeta=1$.
Since $\Lambda \gg 1$, 
$\zeta$ will remain close to one until $2\Lambda(z_1/z)^4\sim 1$.
Thus, $\gamma\propto r$ until $z=z_\gamma$, where
\beq
\frac{z_\gamma}{z_1} \sim \Lambda^{1/4};
\label{am_rg1}
\eeq
note that for $z_1=r_{\rmn{ph}}/r_i$, $r_\gamma/r_i\sim2\times10^3T_i/m_e$
for $\gamma_i$ and $(\siga+\sigs)/\sigt$ of order unity.
In this simplified model, $\gamma\to\gamma_{\infty}$,
the asymptotic Lorentz factor, for $z>z_\gamma$.
Moreover  $\gamma_{\infty}/z_\gamma \sim \gamma_1/z_1$, so
\beq
\gamma_{\infty} \approx \gamma_1 \Lambda^{1/4}.
\label{am3}
\eeq
The resulting `boost' $\gamma_\infty/\gamma_1$ may be large: For 
example if $\gamma_i\sim 2$ and
$\siga+\sigs \sim \sigt$, then $\Lambda \sim 4\times10^6$, and
the asymptotic Lorentz factor of the
flow increases by a factor of $\sim 45$ as compared to its
value at the photosphere!
Specifically, by using previous estimates for the flow variables
at $\tau>1$, we get
\beq
\gamma_{\infty} \sim 1.4 \times 10^3 \gamma_{i}^{3/4} \frac{T_i}{m_e} \left(
\frac{\siga}{\sigt} + \frac{\sigs}{\sigt} \right)^{1/4}
\left( \frac{r_i}{10^6\mbox{cm}} \right)^{1/4}.
\label{am_ginfty}
\eeq

We are now in a position to justify our assertion that the radiation
field will stay close to blackbody in the comoving frame
long after the
optical depth drops below unity. 
In the free streaming limit
the radiation energy, flux and pressure
remain close to their equilibrium values as long as $\zeta-1$ 
is small, which will be
the case until $r \sim r_{\gamma}$. 
For $r>r_{\gamma}, \gamma \to \gamma_{\infty}$, and the right-hand side 
in the pair equation can be neglected. This is because
annihilation becomes unimportant for $r>r_{\gamma}$ because of the low number 
density of pairs, as can be easily checked.
Thus, for $r>r_{\gamma}$, $n_er^2
\sim $ constant and the optical depth at $r_{\gamma}$ becomes 
$\tau(r_\gamma) \equiv \tau_\gamma \approx 
n_{e}(r_{\gamma}) (r_\gamma/\gamma_\infty)(\siga+\sigs)$.
Using our previous results, we find that little
deviation from blackbody radiation 
develops until
$\tau_{\gamma} \sim 1.7\times 10^{-5}\left( \frac{\siga}{\sigt} + 
\frac{\sigs}{\sigt} \right)^{1/4} \gamma_{i}^{3/4}.$

Finally, we need to estimate the asymptotic number density of
pairs. Equation (\ref{am_n}) gives the approximate 
number density 
for $\req<r<r_\gamma$, where $\gamma\propto r$:
\beq
n_e(r) \approx \frac{ (4/11)(\gamma_i/r_i)(\req/r_m)
(1/\pi r_{e}^{2}) }
{1+(1/3)\sqrt{4/11} (\req/r_m) \left[ 1-
\left(\req/r\right)^3 \right]} \left( \frac{\req}{r} \right)^3
\quad \mbox{for $r > \req; \gamma \propto r$.}
\label{am_nb}
\eeq
For $z>z_{\gamma}$ the pair equation reduces to
\beq
n_e(r)z^2 \approx n_e(r_{\gamma})z_{\gamma}^2 \approx n_e(r_{\gamma})z_{\rmn{ph}}^2 \Lambda^{1/2}.
\eeq
Using the approximate results for $n_e(r_{\gamma})$, $r_m/r_i$, $\req/r_m$
and $\rph/\req$, we find
\beq
n_e \sim 8.6 \times 10^{19} \left( \frac{T_i}{m_e} \right)^2 
\left( \gamma_i \right)^{5/4}
\left( \frac{\siga}{\sigt} + \frac{\sigs}{\sigt} \right)^{-1/4} 
\left( \frac{r}{r_i} \right)^{-2}  \mbox{cm}^{-3}
\label{am_nec}
\eeq
for $r>r_{\gamma}$.

The acceleration of the pair wind increases the energy outflow in pairs
to a fraction
$L_e/L \approx { 45 r^2 \gamma^2 \rho_M}/
{ 11\pi^2 r_{i}^{2}\gamma_{i}^{2}T_{i}^{4} }$
of the total luminosity.
Using $\rho_M \approx 2m_e n_e$ and earlier estimates for the other
quantities in the above ratio, we find
\beq
L_e/L \sim 8.5 \times 10^{-6} \left( \gamma_i \right)^{3/4} 
\left( \frac{\siga}{\sigt} + \frac{\sigs}{\sigt} \right)^{1/4}
\label{am_eratio}
\eeq
for $r>r_{\gamma}$
We conclude that for reasonable $\gamma_i$
the flow never becomes matter dominated.
Almost all of the original $L$ is carried away by photons.
Notice that although the asymptotic pair density and Lorentz 
factor depend on both $\gamma_i$ and $T_i$, the value of
$L_e/L$ at $r\gg r_\gamma$ depends only on $\gamma_i$.

To find the asymptotic behavior of the temperature of pairs,
consider the entropy per particle, $Tds=P_M d(1/2n_e)+d(\rho_M/2n_e)$. 
For $T\ll m_e,$ $P_M \approx 2n_e T$ and $\rho_M \approx 2m_e n_e + 3 n_e T$.
Since $n_e \propto r^{-2}$ for large $r$ we have
\beq
Tds \approx \frac{2}{r}Tdr + \frac{3}{2}dT.
\label{seq}
\eeq
Adiabatic cooling, $s$=constant, would result in $T\propto r^{-4/3}$ (as
could also be obtained from $T\propto n^{2/3}$ and $n \propto r^{-2}$).
However, energy is still deposited by photons in the $e^\pm$ flow
even at $r>r_\gamma$; 
the amount of radiative energy input per particle in a time $t_r \sim r/\gamma$
is
\beq
dq = -G^{0}_{0} t_r/2n_e \approx -G^{0}_{0} r/2n_e\gamma.
\eeq
For $r>r_{\gamma}$, $\eo \approx \fo \approx \po$ and energy conservation
implies that $L \approx 4\pi r^2\gamma^2[\eo + \po + 2\fo] \approx 
16\pi r^2\gamma^2\eo$. Consequently,
$-G^{0}_{0} \approx 2n_e \siga \eo \approx 2n_e\siga L/16\pi r^2\gamma^2$.
Since $Tds = dq$, equation (\ref{seq}) takes the form
\beq
\frac{d\ln T}{d\ln r} \approx -\frac{4}{3} + \frac{ \siga L }
{ 24\pi\gamma^3 r T }
\label{neweq1}
\eeq
with solution
\beq
T \approx \frac{T_\gamma r_{\gamma}^{4/3}}{r^{4/3}} +
\frac{\siga L\left(r^{1/3}-r_{\gamma}^{1/3}\right)}
{8\pi\gamma_{\infty}^{3}r^{4/3}}.
\label{am_tinfty}
\eeq

There are two cases to consider, depending on the relative sizes
of the two terms on the right hand side in equation (\ref{neweq1});
their ratio is $\siga L/32\pi\gamma_{\infty}^{3}rT \sim
7.5\times10^2(\siga/\sigt)\gamma_{i}^{-1/4}[(\siga+\sigs)/\sigt]^{-3/4}(r_i/10^6\mbox{cm})$. 
If this ratio is larger than unity, $T\propto r^{-1}$. If the ratio is much 
smaller than unity, $T\propto r^{-4/3}$ until a radius 
$r/r_\gamma \sim (8\pi\gamma_{\infty}^{3}T_\gamma r_\gamma/\siga L)^3 \sim
3.3\times 10^{-11}(\siga/\sigt)^{-3}\gamma_{i}^{3/4}[(\siga+\sigs)/\sigt]^{9/4}(r_i/10^6\mbox{cm})^{-3}$, after which $T\propto r^{-1}$.
This result could also have been obtained from equation (\ref{eeq21_5.1}).

\subsection{Positronium formation?}
\label{possub}

To this point, we have considered $e^\pm$ annihilation in flight only.
At large radii, $e^\pm$ pairs cool considerably, and it is possible
that recombination to positronium occurs, leading ultimately to
destruction of the pairs. Like hydrogen recombination, positronium 
formation would proceed via a free-bound transition to an excited
state, followed by a radiative cascade to lower energy states,
with increasing probability of pair annihilation.
However, in order for this process to be important,
the rate at which positronium is formed must exceed the
expansion rate of the flow.
To estimate the rate of positronium formation, we use the results
from hydrogen recombination with the reduced mass $\approx m_e$ 
replaced by $m_e/2$.

It is straightforward to show that the `gross' recombination 
rate -- ignoring radiative ionization, which leads to a lower
`net' recombination rate -- is small compared to the expansion
rate $\Gamma_{\rmn{exp}}\sim \gamma/r$ at all points outside $r_{\rmn{eq}}$.
Recall that at $r_{\rmn{eq}}$, the electron temperature is 
$T_{\rmn{eq}}\approx 0.052m_e \gg e^4m_e/4 \equiv \chi_{\rmn{pos}}$, the 
ionization potential of positronium.
As long as $T\gg\chi_{\rmn{pos}}$, the recombination coefficient for 
the $n$th excited state of positronium is 
$\Gamma_n=n_e\langle\sigma v\rangle_n$, where 
(e.g. Rybicki \& Lightman 1979)
\beq
\langle\sigma v\rangle_n \approx \frac{32\sqrt{\pi}}{3\sqrt{3}}
\frac{\alpha^3r_{e}^{2}}{n^3}\left(\frac{m_e}{T}\right)^{3/2}
\ln\left(\frac{2Tn^2}{\chi_{\rmn{pos}}}\right).
\eeq
At $r_{\rmn{eq}}$, we find $\Gamma_1/\Gamma_{\rmn{exp}}\sim10^{-3}$; for $n>1$, the 
ratio $\Gamma_n/\Gamma_{\rmn{exp}}$ is even smaller.
Moreover, $\Gamma_n/\Gamma_{\rmn{exp}}\propto n_eT^{-3/2}r/\gamma\propto r^{-3/2}$
outside $r_{\rmn{eq}}$, but still at radii where $\gamma\propto r$, $T\propto 1/r$
and $T>\chi_{\rmn{pos}}$. Thus, $e^\pm$ recombination is very slow in this 
regime.

Ultimately, $T$ drops below $\chi_{\rmn{pos}}$ and the recombination rate
(summed over all states) is approximately
\beq
\Gamma_{\rmn{pos}}\approx 2.08\times10^{-13}n_e\lambda^{1/2}(0.429+(1/2)\ln\lambda
+0.469\lambda^{-1/3})
\eeq
where $\lambda=1.33\times10^{-5}m_e/T$ and $n_e$ is in $\mbox{cm}^{-3}$ 
(e.g. Rybicki \& Lightman 1979). In this regime, $\Gamma_{\rmn{pos}}/\Gamma_{\rmn{exp}}
\propto n_er/\gamma T^{1/2}\propto 1/r\gamma^2T^{1/2}$, since
$n_er^2\gamma\approx$ constant; thus, $\Gamma_{\rmn{pos}}/\Gamma_{\rmn{exp}}$
continues to decrease with radius far out in the flow, at first
$\propto r^{-5/2}$, and ultimately $\propto r^{-1/2}$. Thus we 
conclude that the formation of positronium is always very slow,
and does not affect the pair density substantially.

\section{Results}
\label{results}

A large number of models were calculated, covering a wide range
of initial temperatures ($1\le T_i/m_e\le 1000$), initial 
Lorentz factors ($1.25 \le \gamma_i \le 10$), and absorption
cross-sections ($10^{-6} \le \siga/\sigt \le 1$).
The scattering cross-section was held fixed at $\sigs/\sigt = 1$, and
the initial radius $r_i=10^6$cm.
Due to the stiffness in the pair equation, and in the dynamical 
equations for $\tau < 1$, we used a semi-implicit scheme for 
solving the system of equations.

\begin{figure}
\begin{picture}(450,450)(-25,0)
\put(0,3){\epsfxsize=6.0in\epsffile{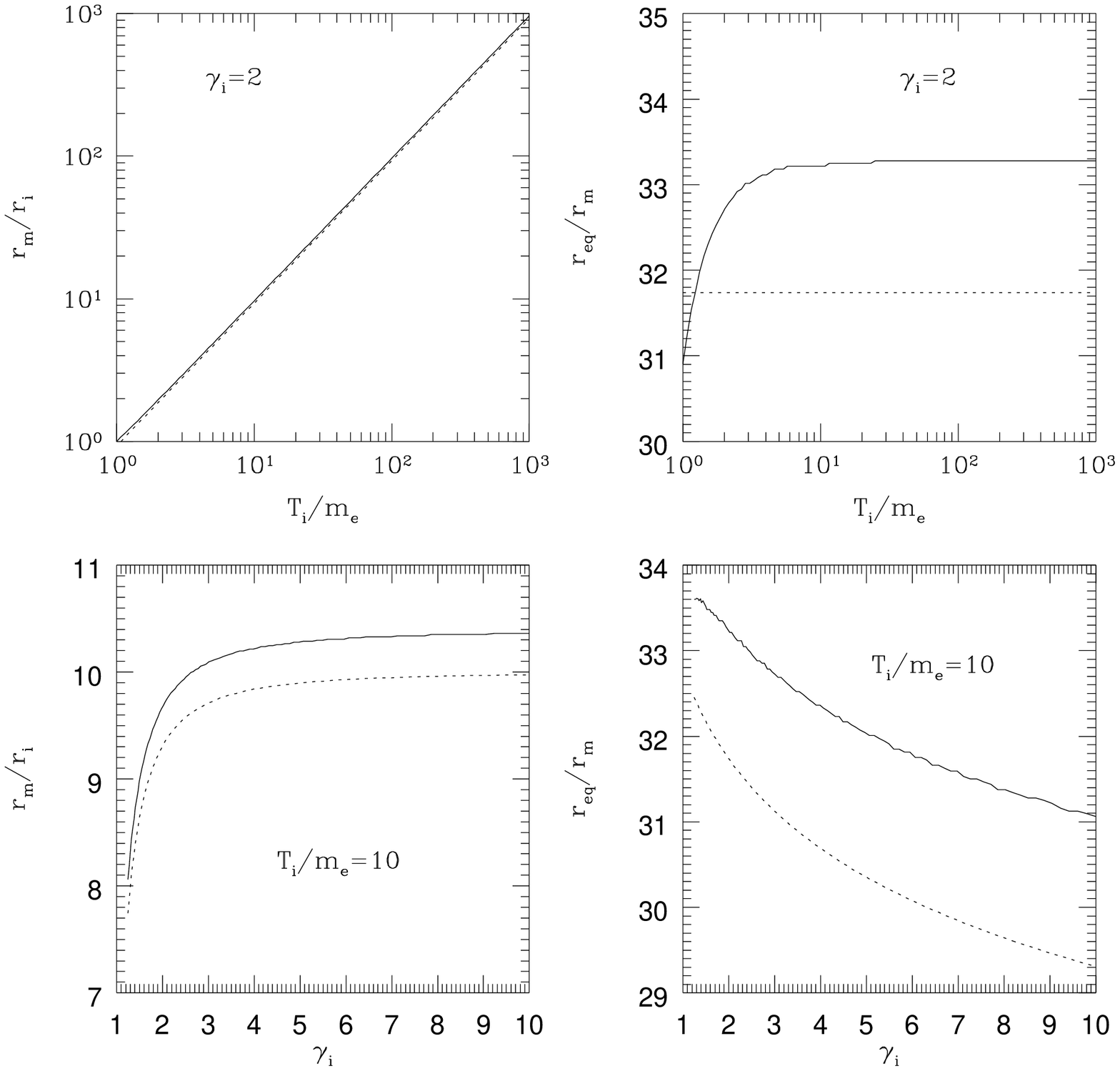}}
\end{picture}
\caption{The ratios $r_m/r_i$ and $\req/r_m$ 
are shown as functions of the initial temperature $T_i$ for
fixed initial Lorentz factor $\gamma_i$ in the upper panels,
and as a function of $\gamma_i$ for fixed $T_i$ in the lower
panels. $r_i$ is the initial radius, $r_m\equiv r(T=m_e)$ and
$\req \equiv r(n_e=2n_{e,\rmn{eq}})$. The absorption cross section
is $\siga=10^{-3}\sigt$. The solid lines correspond to the 
numerical solution, and the dotted lines to the analytical model.  
\label{fig1}}
\end{figure}

\begin{figure}
\begin{picture}(450,450)(-25,0)
\put(0,3){\epsfxsize=6.0in\epsffile{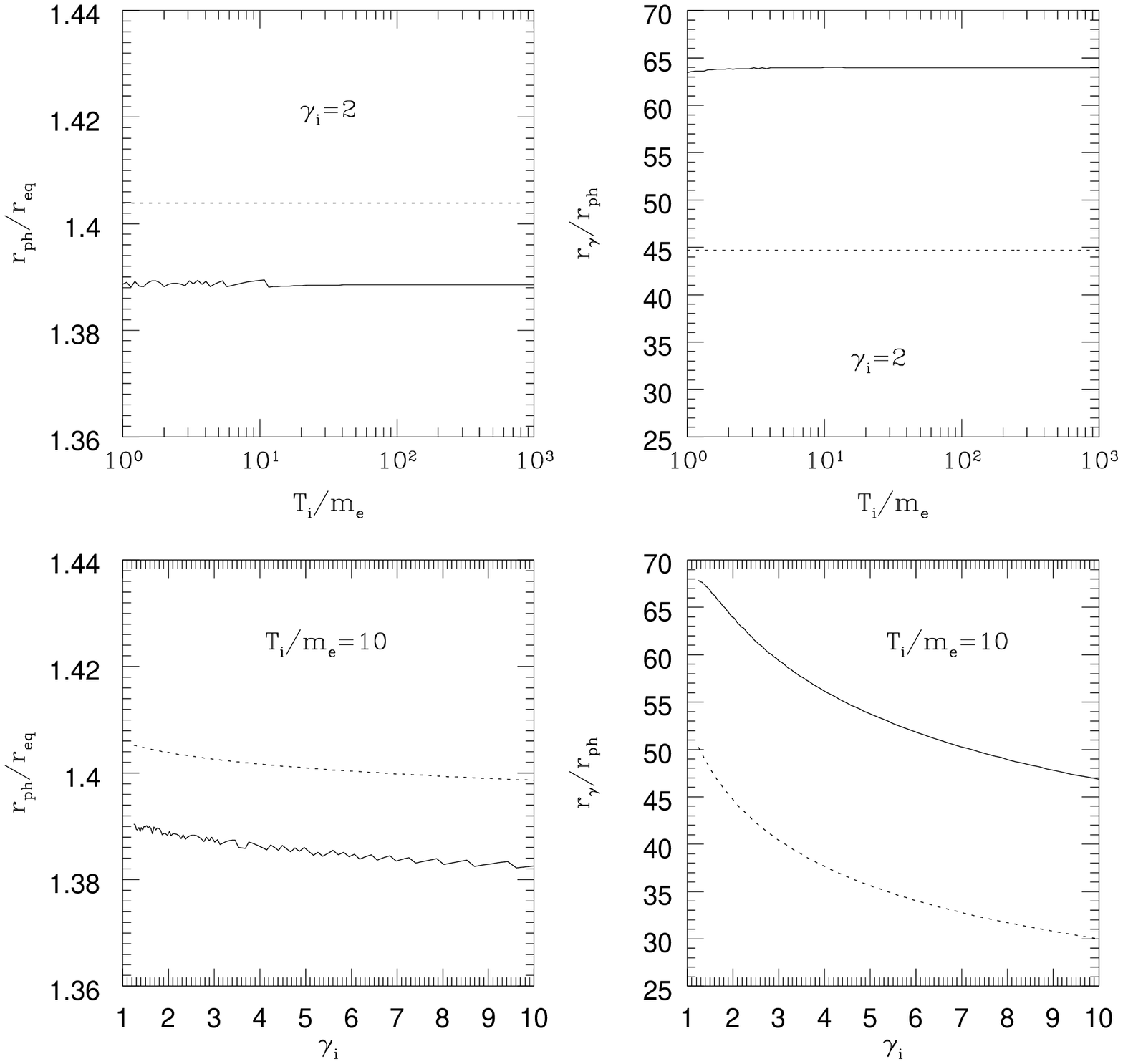}}
\end{picture}
\caption{The ratios $\rph/\req$ and $r_\gamma/\rph$ are 
shown as functions of initial temperature $T_i$ and initial 
Lorentz factor $\gamma_i$. $\req \equiv r(n_e=2n_{e,\rmn{eq}})$,
$\rph \equiv r(\tau=1)$ and $r_\gamma \equiv r(\zeta=1/2)$.
The absorption cross-section is fixed at $\siga=10^{-3}\sigt$.
The solid lines correspond to the numerical solution, and 
the dotted lines to the analytical model.   
\label{fig2}}
\end{figure}

Figs \ref{fig1} and \ref{fig2} show the ratios $r_m/r_i$, 
$\req/r_m$, $\rph/\req$ and $r_\gamma/\rph$ as functions of the 
initial temperature $T_i$ and Lorentz factor $\gamma_i$. The 
agreement between the numerical results (solid lines) and the
approximations from the analytical model (dotted lines) is
very good.

\begin{figure}
\begin{picture}(200,225)(-130,0)
\put(0,3){\epsfxsize=3.0in\epsffile{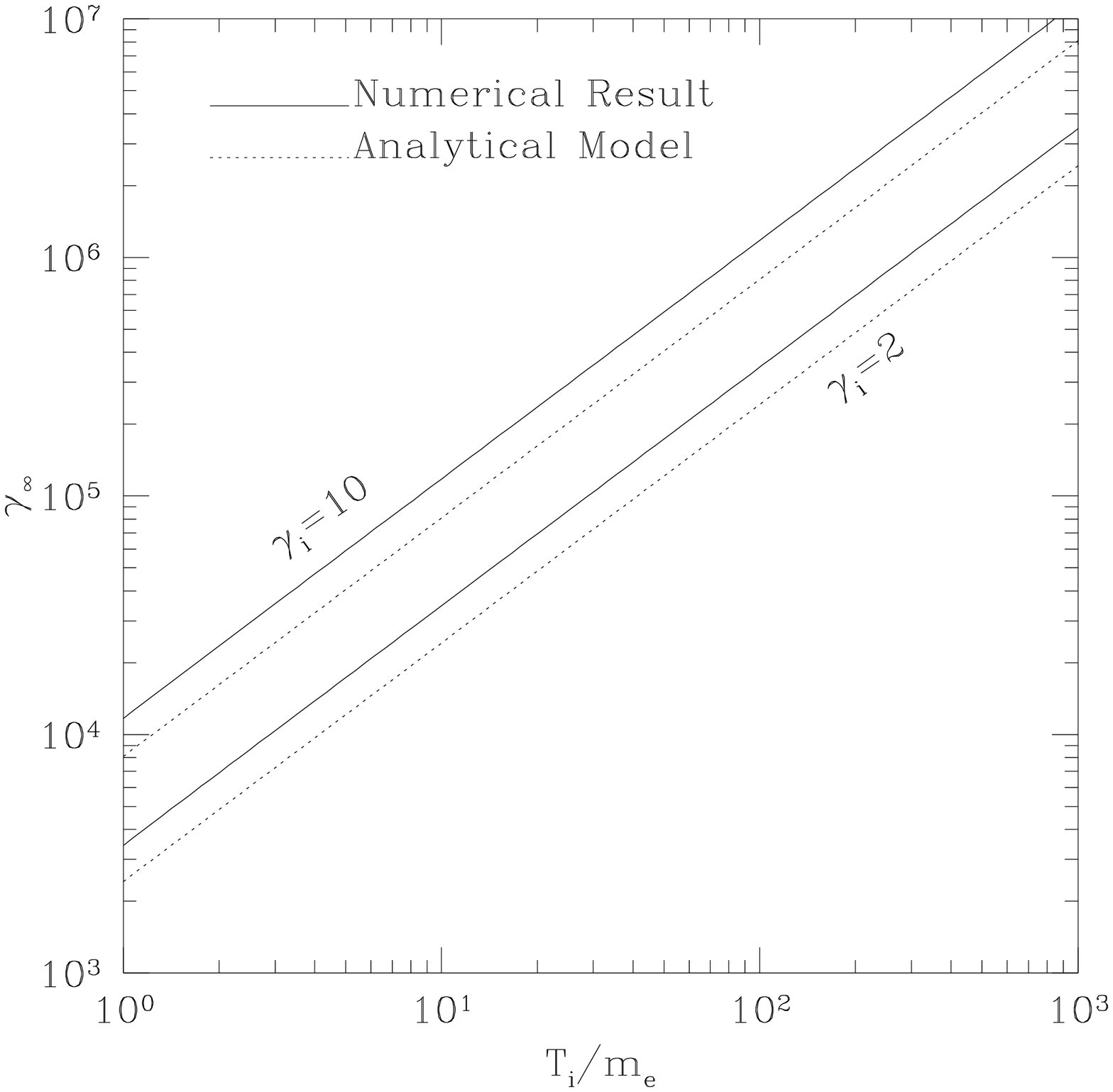}}
\end{picture}
\caption{The asymptotic Lorentz factor 
as a function of the initial temperature for $\gamma_i=2$ and
$\gamma_i=10$. The absorption cross-section is held fixed at
$\siga = 10^{-3}\sigt$. 
\label{fig3}}
\end{figure}

Fig. \ref{fig3} confirms
the high asymptotic Lorentz factors as well as their dependence
on initial temperature and Lorentz factor, $\gamma_\infty \propto
(T_i/m_e)\gamma_{i}^{3/4}$. 

\begin{figure}
\begin{picture}(350,375)(-50,0)
\put(0,3){\epsfxsize=5.0in\epsffile{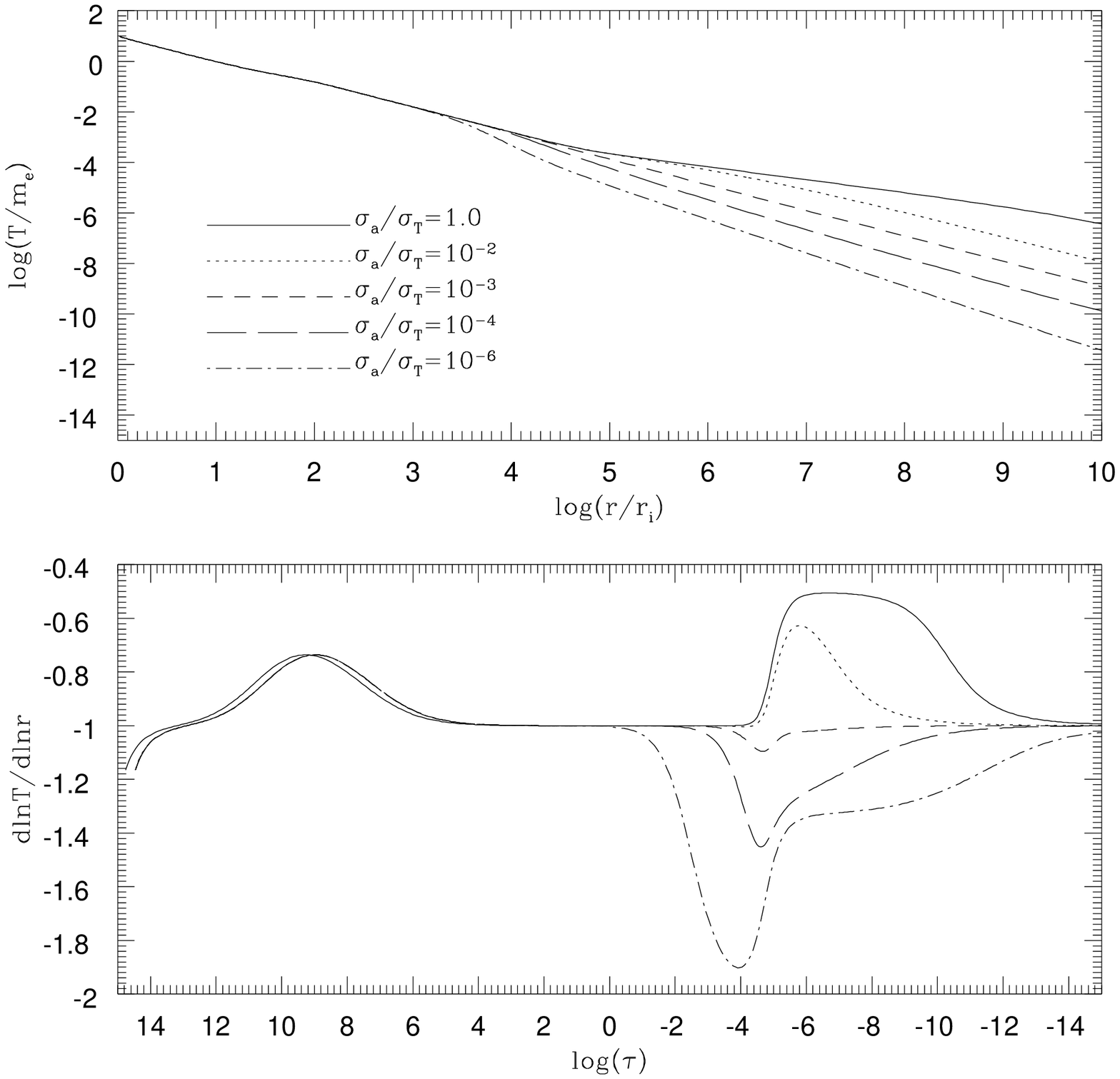}}
\end{picture}
\caption{The local temperature $T(r)$ in units of $m_e$, and
its derivative $d\ln T/d\ln r(\tau)$, are plotted for different
absorption cross-sections. In these plots, $T_i=10m_e$ and $\gamma_i=2$.
\label{fig4}}
\end{figure}

Equation (\ref{am_tinfty}) implies that $T\propto 1/r$ for large $r$;
this is seen in Fig. {\ref{fig4}}.
The behavior of $d\ln T/d\ln r$ can be understood from the 
dynamical equations $T^{\alpha \beta}_{;\beta} = -G^\alpha$. For $T \ll m_e$, 
$\rho_M \gg P_M$ and the energy equation can be approximated as
\beq
\frac{3}{2} \frac{d\ln T}{d\ln r} + 2 + \frac{d\ln \gamma}{d\ln r} 
\approx \frac{z}{\gamma} (\sigt r_i m_{e}^{3})\frac{m_e}{T} 
\frac{\siga}{\sigt} \left[ \frac{E_0}{m_{e}^{4}} - 
\frac{U_{\rmn{eq}}}{m_{e}^{4}} \right].
\label{r_1}
\eeq
The left hand side of this equation contains terms of order unity.
If $\siga/\sigt$ is small enough to make the right hand side much 
less than 1, we must have $d\ln T/d\ln r \approx 
-4/3 - (2/3)d\ln \gamma/d\ln r$. 
For $r<r_\gamma$, where $\gamma \propto r$, $d\ln T/d\ln r$ will 
therefore tend to approach $-2$, as can be seen from Fig. {\ref{fig4}} when
$\siga/\sigt = 10^{-6}$. When $r>r_\gamma$, $\gamma \to$ constant
and $d\ln T/d\ln r$ approaches $-4/3$. This is also seen 
in the graph, before heating from the radiation eventually
forces $T$ to be proportional to $1/r$.
On the other hand, if $\siga/\sigt$ is large enough to make
$(z/\gamma)(\sigt r_i m_{e}^{3})(m_e/T)(\siga/\sigt) \gg 1$, 
$E_0/m_{e}^{4} - U_{\rmn{eq}}/m_{e}^{4}$ must be very close to 0.
From the free-streaming form of $E_0$ this means that $T^4 \propto 
r^{-2}\gamma^{-2}$. Consequently, for $r<r_\gamma$ we have 
$d\ln T/d\ln r \approx -1$, while for $r>r_\gamma$, $\gamma \sim
\gamma_\infty$ and $T\propto r^{-1/2}$. This is seen from the 
plot of $d\ln T/d\ln r$ for $\siga=\sigt$.

\begin{figure}
\begin{picture}(450,450)(-25,0)
\put(0,3){\epsfxsize=6.0in\epsffile{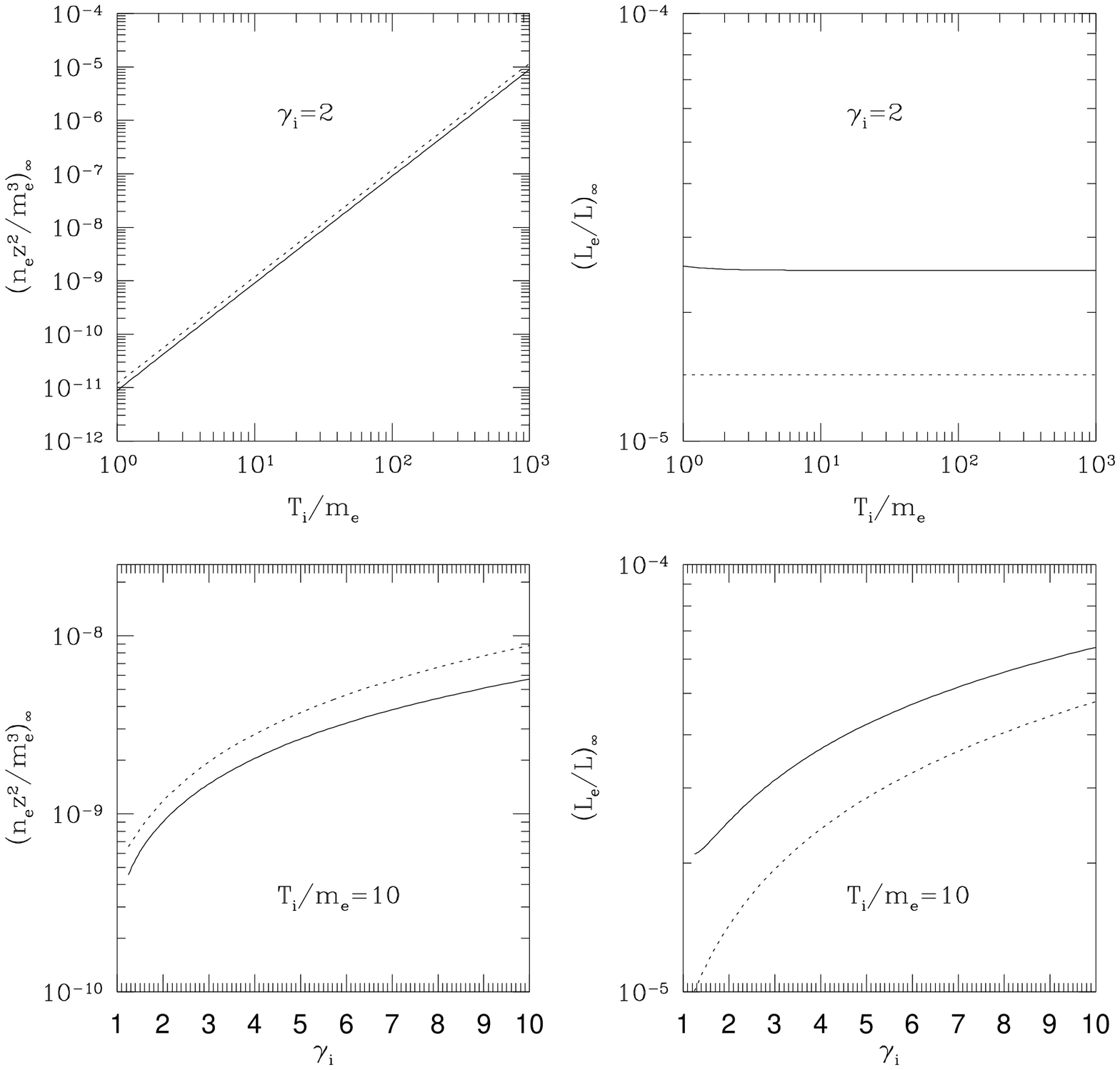}}
\end{picture}
\caption{The asymptotic pair number density $n_e$ times
$z^2=(r/r_i)^2$ is shown in the left panels in units
of $m_{e}^{3}$ as a function of initial temperature $T_i$ 
and Lorentz factor $\gamma_i$. The right panels show the
asymptotic energy content in the pairs relative to that 
in the radiation. In all plots, $\siga=10^{-3}\sigt$.  
Solid lines represent the numerical solution and dotted lines the 
analytical model.
\label{fig5}}
\end{figure}

Next, Fig. \ref{fig5} serves as a check of our estimates for the
asymptotic number density of pairs, and for the ratio of energy in pairs 
to the total energy. Again, the analytical model (dotted lines) agrees 
very well with the numerical results (solid lines).

\begin{figure}
\begin{picture}(350,375)(-50,0)
\put(0,3){\epsfxsize=5.0in\epsffile{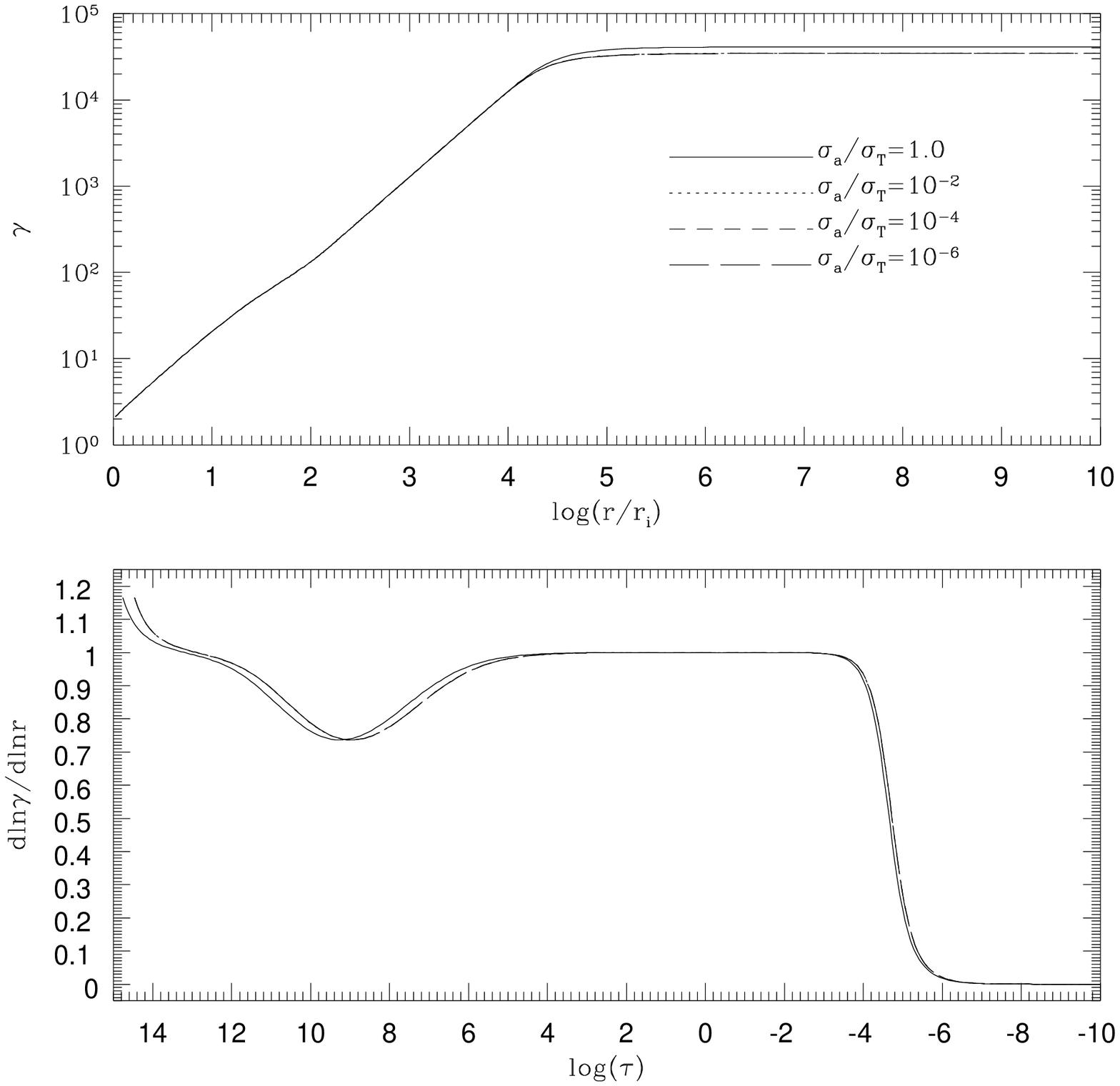}}
\end{picture}
\caption{The Lorentz factor $\gamma(r)$ and its derivative
$d\ln \gamma/d\ln r(\tau)$ are shown for a set of different
absorption cross-sections. $T_i=10m_e$ and $\gamma_i=2$ are
fixed.  
\label{fig6}}
\end{figure}

When $\tau < \tau_\gamma$, i.e. $r>r_\gamma$, the Lorentz factor 
approaches a constant value, as seen in Fig. \ref{fig6}. The 
dip in $d\ln \gamma/d\ln r$ for large optical depths happens
when $T\sim m_e$, after which pairs contribute little to the 
total energy. This point is the transition between $\gamma 
\approx \gamma_i (r/r_i)$, when $T\gg m_e$, and the low-temperature
limit $\gamma \approx \gamma_i (r/r_i)\sqrt{4/11}$ (see equations 
[\ref{am_gshift1}] and [\ref{am_gshift2}]).

\section{Robustness of Blackbody}
\label{spec}

Here, we show that the deviations from blackbody in the photon 
distribution function are small. 
In the $\gamma\gg 1$ limit, the corresponding 
spectrum observed
in the lab frame has a slightly broader peak and
a shallower slope for low frequencies compared to a blackbody. 
We also consider non-radial perturbations to the flow and
resulting effects on the observed radiation spectrum.

\subsection{Deviations from blackbody in the comoving frame}

In this subsection we will give a more careful justification for the simple model 
we used for the radiation field.
When the optical depth is large,
the radiation is trapped effectively. Because thermal emission and
scattering are isotropic in the rest frame of the flow, one 
expects the radiation field to be close to isotropic when the 
optical depth is larger than one. The deviation from isotropy can 
be treated as a small perturbation on the radiation field in 
this limit.
From Section \ref{sub_tault1}, the equation of radiative transfer is
\beq
k^{\beta} \frac{\pl}{\pl x^{\beta}}\fg(k,x) = C_{\gamma}.
\eeq
The Lorentz invariant collision terms are
\beq
{\cal C}_s=2n_e\sigma_s \tilde k^0 \left[-\fg(k,x)
+\int{d^2\hat n^\prime\,g(\hat n^\prime\to\hat n)\fg(k^\prime,x)}\right]
\qquad \mbox{and}
\eeq
\beq
{\cal C}_a=2n_e\sigma_a \tilde k^0[-\fg(k,x)+\fbb(\tilde k^0/T)]
\eeq
for scattering and absorption/emission, respectively.
Below, we neglect pair production and annihilation since these
processes have a negligible effect on the escaping radiation.

Perturbing away from blackbody radiation, we write
\beq
\fg(k,x) = \fbb(\ktilo/T) + \delta \fg(k,x).
\label{d_2}	
\eeq
It is convenient to introduce the four-vector
$n^\mu = k_\nu P^{\mu\nu}/\tilde k^0$,
where the projection tensor is
$P^{\mu\nu}=g^{\mu\nu}+U^\mu U^\nu.$
This four-vector reduces to the unit vector along the direction 
of propagation in the rest frame of the flow. 
In terms of $n^\mu$ and $\tilde{k}^0=-k_\mu U^\mu$, we have 
$k^\mu = \tilde k^0 (n^\mu + U^\mu )$.
 
Now expand the correction to blackbody, $\delta \fg$, in Legendre 
polynomials:
\beq
\delta \fg(k,x)=A+B_\mu n^\mu+
C_{\mu\nu} \left( {3n^\mu n^\nu - P^{\mu\nu}\over 2}
\right).
\label{d_3}	
\eeq
Introduce an explicit form for
the scattering kernel,
\beq
g(\nhatprime\to\nhat)={1+aP_2(\nhat\cdot\nhatprime)
\over 4\pi},
\eeq
where $P_2(\hat n, \hat n^\prime) = [3(\hat n \cdot \hat n^{\prime})^2 -1]/2$.
We will specialize eventually to Thomson scattering for which $a=1/2$.

With the expansions in equations (\ref{d_2}) and (\ref{d_3}), the
right-hand side of the transfer equation becomes
\beqa
C_\gamma & = & 2n_e\ktilo \left\{-\sigma_a A
-(\sigma_a+\sigma_s)B_\mu
n^\mu  \right.   \nonumber  \\
 & - & \left.  \left[\sigma_a+\sigma_s\left(1-{a\over 5}\right)
\right]C_{\mu\nu}\left({3n^\mu n^\nu-P^{\mu\nu}\over 2}
\right)\right\} + C_p. 
\label{d_5}
\eeqa
On the left hand side one gets
\beqa
 & k^\beta \frac{\partial}{\partial x^\beta} \fbb(\ktilo/T(x)) = \nonumber \\
 & -\ktilo\psi(\ktilo/T)\left[
U^\nu\frac{T_,\nu}{T} + \frac{P^{\mu\nu}}{3}U_{\mu;\nu} +
n^\mu\left(\frac{T_{,\mu}}{T} + U^\nu U_{\mu;\nu}\right) +
\frac{2}{3}\left(\frac{3n^\mu n^\nu-P^{\mu\nu}}{2}\right)U_{\mu;\nu}\right]
\label{d_6}
\eeqa
with $\psi(z)\equiv z \fbb^{\prime}(z)$.
Equating equations (\ref{d_5}) and (\ref{d_6}), the results 
for $A, B_\mu$ and $C_{\mu \nu}$ can be read off.
This allows us to find the 
energy momentum tensor for radiation,
\beq
T_{R}^{\mu\nu} = \int{{d^3k\over k^0}k^\mu k^\nu
\left[\fbb(-k^\lambda U_\lambda/T(x)) + A +  B_\lambda n^\lambda+
C_{\lambda\sigma} \left({3n^\lambda n^\sigma - 
P^{\lambda\sigma}\over 2}
\right)\right]}.
\label{d_7}
\eeq
The results are
\beqa
\delta \tilde E_0 & \equiv & \frac{E_0-U_{\rmn{eq}}}{U_{\rmn{eq}}} =
-4t_0 \left\{ \frac{1}{\gamma T} \frac{d}{d\ln r} (T \sqrt{\gamma^2 -1} ) 
- \frac{2}{3} \frac{r}{\gamma} \frac{d}{d\ln r} \left( \frac{\sqrt{\gamma^2-1}}{r} \right) \right\}   \nonumber  \\
 & + &   t_0\frac{r}{\gamma}G_{\rmn{p,0}}^{0}\frac{1}{U_{\rmn{eq}}}
\label{d_12}   \\
\delta \tilde F_0 & \equiv & \frac{F_0}{U_{\rmn{eq}}} = 
- \frac{4}{3} t_1 \frac{1}{\gamma T} \frac{d}{d\ln r} (\gamma T)
   \label{d_13}   \\
\delta \tilde P_0 & \equiv & \frac{P_0 - U_{\rmn{eq}}/3}{U_{\rmn{eq}}} =
\frac{\delta \tilde E_0}{3} - \frac{16}{45} t_2 \frac{r}{\gamma} 
\frac{d}{d\ln r} \left( \frac{\sqrt{\gamma^2 - 1}}{r} \right).
\label{d_14} 
\eeqa
where
\beqa
t_0 & \equiv &
-\frac{\gamma}{r}{\pi\over 2n_eaT^4}\int_0^\infty{d\ktilo(\ktilo)^3\psi(\ktilo/T)\over\sigma_a}   \\
t_1 & \equiv &
-\frac{\gamma}{r}{\pi\over 2n_eaT^4}\int_0^\infty{d\ktilo(\ktilo)^3\psi(\ktilo/T)\over\sigma_a+
\sigma_s}    \\
t_2 & \equiv &
-\frac{\gamma}{r}{\pi\over 2n_eaT^4}\int_0^\infty{d\ktilo(\ktilo)^3\psi(\ktilo/T)\over\sigma_a+
\sigma_s(1-a/5)}   
\eeqa
with $U_{\rmn eq}=\pi^2T^4/15$ 
as before.

From the analytical model as well as the numerical results we know that 
$\gamma T$ = constant and $\gamma \propto r$ are
excellent approximations for $r<r_\gamma$.
Then $\delta \tilde F_0 = 0$, $\delta \tilde E_0 \sim
{\cal O}(t_0/\gamma^2)$ and $\delta \tilde P_0 \sim {\cal O}(t_2/\gamma^2)$
when $\gamma \gg 1$; since $t_{0}^{-1}$ and $t_{2}^{-1}$ are
of the same order as the absorption and total optical depth, 
respectively, we see that the corrections to 
blackbody radiation are negligible when $\gamma \gg 1$.

These results are also consistent with a more careful treatment
of the equation of radiative transfer. This equation is most conveniently written as
\beqa
\left[\parr
+{2(1-\lambda)\over r}\left({r\over \gamma}{d\gamma\over dr}-1\right)
\left(q\parq
-\lambda\parlam\right)\right]\ngamqlamr =   \nonumber  \\
{\ne (\siga + \sigs) \over\gamma\lambda}
\left[-\ngamqlamr + \epsilon \nbbq+
(1-\epsilon)
\int_0^{+1}{d\lamp\ngamqlamrp}\right]  \label{d_radtransf}
\eeqa
in an extremely relativistic flow (Grimsrud 1998).
Here $\epsilon \equiv \siga/(\siga+\sigs)$,
$\eta \equiv 2\gamma^2 (1-\mu)$, $\lambda \equiv 1/(1+\eta)$ and
$q \equiv k/2\lambda$, with $\mu$ the direction cosine and $k$ 
the photon energy in the lab frame. Again, the contribution from pairs
is neglected on the right hand side of the transfer equation.
The left hand side of this equation simplifies considerably when 
$\gamma \propto r$. If, in addition, $\gamma T$ = constant, we see
that $\fg(q,\lambda,r) = \fbb(q/\gamma T)$ is the solution
for the distribution function. Thus, as long as these two conditions are
satisfied, the deviation from blackbody radiation vanishes in the
high-$\gamma$ limit up to 
terms of order ${\cal O}(1/\gamma^2)$ which we have neglected.
Note that equation (\ref{d_radtransf}) holds for both large and 
small optical depths.

It is interesting to contrast the high--$\gamma$ limit of equations
(\ref{d_12})-(\ref{d_14}) with a static atmosphere, where $\gamma = 1$.
In the static case, neglecting pair-heating, 
\beq
\delta \tilde E_0 = 0, \qquad \delta \tilde F_0 = -\frac{4}{3}t_1 
\frac{d\ln T}{d\ln r}, \qquad \mbox{and} \qquad \delta \tilde P_0 = 0
\eeq
in the diffusion approximation, $\tau\gg1$. 
Often one relates the radiation pressure to the radiation energy via
a variable Eddington factor $f_{\rmn{Edd}}(\tau)$, $P_0=E_0[1/3 + f_{\rmn{Edd}}(\tau)].$
This is a closure relation needed
for the first two moment equations obtained 
from the equation of radiative transfer.
The above relations constrain the form of
$f_{\rmn{Edd}}(\tau)$. For $\tau \gg 1$, $f_{\rmn{Edd}}(\tau)$ 
must be at least quadratic in 
$1/\tau$; the linear term vanishes in this limit.

Recall from Section \ref{am_tault1} that a free-streaming radiation
field, coupled to the matter via the dynamical equations 
$T_{M; \beta}^{\alpha \beta} = -G^\alpha$, resulted in $\gamma \propto r$ 
and consequently a radiation field close to blackbody out to 
optical depths much less than 1. This behavior is then also 
consistent with the full radiative transfer equation as
discussed above.
Furthermore, note that since deviations from blackbody in the 
radiation fields start building up only when $\tau \ll 1$, 
we cannot use a variable Eddington factor $f_{\rmn{Edd}}(\tau)$ in our 
extremely relativistic atmosphere.

\subsection{Observed spectrum}
\label{obs_spec}

\begin{figure}
\begin{picture}(200,225)(-125,0)
\put(0,3){\epsfxsize=3.0in\epsffile{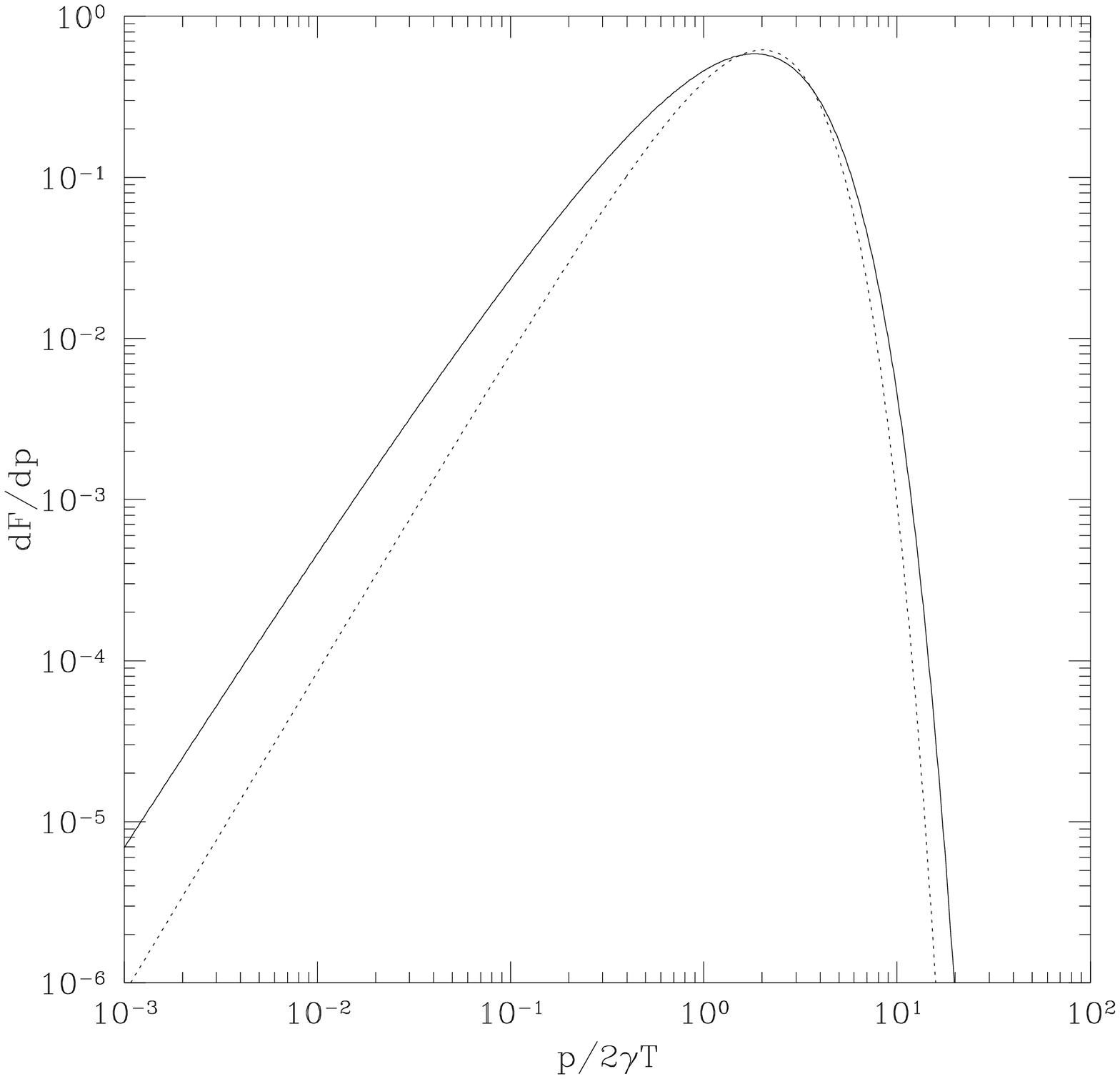}}
\end{picture}
\caption{The spectrum seen by an observer in 
the lab frame (solid line) compared to blackbody (dotted line).
The units for the flux are arbitrary.
\label{spec_fig}}
\end{figure}

To compute the observed spectrum, we use the fact 
that the (comoving frame) photon distribution function
is blackbody to a very good approximation at any $\tau$ for $\gamma\propto r$.
This allows us to find
the radially directed flux per energy interval seen by a stationary
observer at arbitrary $r$, using
\beq
\frac{dF}{dp} = \frac{p^3}{2\pi^2}\int_{-1}^{1}\frac{\mu d\mu}
{\exp[\gamma p(1-v\mu)/T]-1}
\label{spec_eq}
\eeq
where $\gamma\propto r$ (and $p$ is the observer--frame photon energy)
and $T$ is the 
blackbody temperature in the frame moving radially with Lorentz factor 
$\gamma$. (The same results can be found from free streaming at $\tau<1$.)
For $\gamma\gg1$, the integral can be done analytically to give
\beq
\frac{dF}{dp} \approx \frac{p^2T}{2\pi^2\gamma}\{
-\ln[1-\exp(-p/2\gamma T)]\}.
\label{seq1}
\eeq
At small values of $p/\gamma T$, this expression tends to
\beq
\frac{dF}{dp} \approx \frac{p^2T}{2\pi^2\gamma}\ln(2\gamma T/p);
\eeq
at large values of $p/\gamma T$ it tends to
\beq
\frac{dF}{dp} \approx \frac{p^2T}{2\pi^2\gamma}\exp(-p/2\gamma T).
\eeq
In order to compare (\ref{seq1}) with a blackbody that fits the 
observed flux best, we minimize the integral 
\beq 
S(B_0,x_0) = \int_{0}^{\infty}dx\left\{ 
-x^2\ln[1-\exp(-x)] - \frac{B_0x^3}{\exp(x/x_0)-1} \right\}^2
\eeq
with respect to the two parameters $B_0$ and $x_0$.
The two functions are plotted in Fig. \ref{spec_fig}.
Compared to a blackbody, the observed flux is seen to have a 
broader peak, and a shallower slope at low photon energies.
A similar result was found by Goodman (1986) from his fireball simulations.

\subsection{Perturbations about spherical symmetry}

\label{pert}

So far we have assumed the flow to be static and spherically symmetric.
In reality, however, these assumptions break down if the flow is 
affected by, e.g., a magnetic field, an anisotropic and/or variable 
temperature distribution at the inner surface $r=r_i$, or inhomogeneities
in the external medium. Here we will merely look at the effects on a static
flow due to perturbations about its spherically symmetric solution.

For simplicity, we assume the energy density and pressure to be
functions of temperature only, i.e. we use the equilibrium expression 
for the number density of electrons and positrons. As we have seen, this
is a good approximation inside the photosphere as far as the dynamics of 
the flow is concerned.
Also, the flow is 
assumed to be azimuthally symmetric.

Defining the quantity $\sigma \equiv (\rho+P)/T$, the energy-momentum tensor can be written as
\beq
\talbe = T\sigma U^\alpha U^\beta + P\galbe.
\eeq
Contracting the equations of motion ($\talbe_{;\beta}=0$) with $U_\alpha$ 
results in the conservation law
\beq
\left( \sigma U^\mu \right)_{;\mu} = 0.
\label{prt2}
\eeq
This can be used to simplify the equations of motion;
\beq
\left( \galbe + \ual \ube \right)T_{,\beta} + T \ube \ual_{;\beta} = 0.
\label{prt3}
\eeq
We now introduce the four-vector
$V^\alpha = TU^\alpha$,
which allows us to write equation (\ref{prt3}) as
\beq
V^\alpha\left( V_{\alpha;\beta} - V_{\beta;\alpha} \right) 
\equiv V^\alpha \oalbe = 0.
\eeq
We have here defined the vorticity tensor
$\oalbe \equiv \left( \tual \right)_{;\beta} - \left( \tube \right)_{;\alpha}=
\left( \tual \right)_{,\beta} - \left( \tube \right)_{,\alpha}.$
It is straightforward to show that the circulation
\beq
\Gamma \equiv \oint \tual dx^\alpha =
\int \left[ \left( \tual \right)_{;\beta} - \left( \tube \right)_{;\alpha} \right] dx^\alpha dx^\beta =
\int \oalbe dx^\alpha dx^\beta
\label{prt6}
\eeq
is unchanged along the path of a fluid element.

We are interested in axisymmetric ($\partial/\partial\phi=0$) perturbations
around spherically symmetric flow. Let the temperature of the unperturbed
flow be $\Tbar(r)$, and the nonzero components of the unperturbed 
four-velocity be
$\Ubar^0(r)=\gbar(r)$ and $\Ubar^r(r)=\gbar(r)\vbar(r)$; consequently
$\Vbar^0(r)=\Tbar(r)\Ubar^0(r)$ and $\Vbar^r(r)=\Tbar(r)\Ubar^r(r)$, and
$\omega_{\alpha\beta}=0$ for the unperturbed flow. Denote perturbed 
quantities by $\delta T(r,\theta)$, $\delta U^\mu(r,\theta)$ and
$\delta V^\mu(r,\theta)$.
From the fluid equations
$V^\nu\omega_{\mu\nu}=0$ we find, to first order in the perturbations,
\beq
\Vbar^0\omega_{\mu 0}+\Vbar^r\omega_{\mu r}=0.
\label{eq:fl1}
\eeq
The $\mu=0$ and $\mu=r$ components of this equation both imply $\omega_{0r}=
\partial \delta V_0/\partial r=0$, so that $\delta V_0=f(\theta)$, independent of $r$. 
Below, it will prove convenient to expand
\beq
f(\theta)={\Vbar^0\over 2}\sum_{\ell\neq 0}\ellp\tl\pell(\theta),
\label{eq:fthetadef}
\eeq
where $\pell(\theta)$ is the Legendre function. We omit $\ell=0$ in the sum
because it may be absorbed into the background solution, for which 
$\Vbar_0=$ constant. The $\mu=\theta$ component of equation (\ref{eq:fl1})
becomes
\beq
\omega_{\theta r}={\partial\delta V_\theta\over\partial r}
-{\partial\delta V_r\over\partial \theta}
={\Vbar^0\over\Vbar^r}{df(\theta)\over d\theta}
={(\Vbar^0)^2\over 2\Vbar^r}\sum_{\ell\neq 0}
\ellp\tl{d\pell(\theta)\over d\theta}
\label{eq:omtr}
\eeq
after substituting for $\delta V_0$.
The $\mu=\phi$ component of equation (\ref{eq:fl1})
implies $\partial \delta V_\phi/\partial r=0$, so that $\delta V_\phi=g(\theta)$, or
$\sin^2\theta U^\phi=g(\theta)/\Tbar(r)r^2$.

Next, we perturb the entropy equation, which may be written in the form
\beq
{1\over r^2\sin\theta}{\partial(r^2\sin\theta Q(T)V^\mu)\over\partial x^\mu}=0,
\label{eq:entropy1}
\eeq
where $Q(T)\equiv\sigma(T)/T$. To first order we find
\beq
\Vbar^r{\partial\over\partial r}\left(\nu(\Tbar){\delta T\over\Tbar}+
{\delta V^r\over\Vbar^r}\right)
+{1\over r^2\sin\theta}{\partial(\sin\theta\delta V_\theta)\over\partial\theta}
=0
\label{eq:entropy2}
\eeq
where $\nu(T)\equiv d\ln Q(T)/d\ln T$, and $\delta V^\theta=
g^{\theta\theta}\delta V_\theta=r^{-2}\delta V_{\theta}$.

The normalization condition $-T^2=V^\mu V_\mu$ implies, to first order in
the perturbations,
\beq
\Tbar\delta T=-\Vbar^0f(\theta)-\Vbar^r\delta V^r;
\label{eq:norm}
\eeq
using this relationship to eliminate $\delta V^r=\delta V_r$ in equations
(\ref{eq:omtr}) and (\ref{eq:entropy2}) implies
\beq
{\partial(\sin\theta\delta V_\theta)\over\partial r}
+{\Tbar^2\over\Vbar^r}\sin\theta{\partial(\delta T/\Tbar)\over\partial
\theta}=0
\label{eq:omtr2}
\eeq
(where we have also multiplied by $\sin\theta$) and
\beq
{\partial(\sin\theta\delta V_\theta)\over\partial\theta}
+r^2\sin\theta\Vbar^r{\partial[\nu_R(r)(\delta T/\Tbar)]\over
\partial r}={2r\sin\theta f(\theta)\Tbar^2\Vbar^0\over (\Vbar^r)^3}
{d\ln\Tbar\over d\ln r},
\label{eq:entropy3}
\eeq
where $\nu_R(r)\equiv\nu(\Tbar)-(\Tbar/\Vbar^r)^2$. Differentiate
equation (\ref{eq:omtr2}) with respect to $\theta$ and equation 
(\ref{eq:entropy3}) with respect to $r$ to find
\beq
{\Vbar^r\over\Tbar^2}{\partial\over\partial r}
\left[r^2\Vbar^r{\partial\over\partial r}\left(\nu_R(r)
{\delta T\over\Tbar}\right)\right]-{1\over\sin\theta}
{\partial\over\partial\theta}\left[\sin\theta{\partial\over
\partial\theta}\left({\delta T\over\Tbar}\right)\right]=
{2f(\theta)\over\Vbar^0}\Hbar(r),
\label{eq:teqn}
\eeq
where (recall that $d\Vbar^0/dr=0$)
\beq
\Hbar(r)\equiv {\Vbar^r(\Vbar^0)^2\over\Tbar^2}\left[{r\Tbar^2
\over (\Vbar^r)^3}{d\ln\Tbar\over d\ln r}\right]_{,r}.
\eeq

When the background flow is extremely relativistic and dominated
by extremely relativistic particles, we may take $\Vbar^r=
\Vbar^0$, $\Tbar\propto 1/r$, and $\nu_R=2$ up to corrections 
$\sim\gbar^{-2}$. With these substitutions, $\Hbar(r)=1$ and
equation (\ref{eq:teqn}) becomes
\beq
2\gbar^2\frac{\partial}{\partial r}
\left[r^2\frac{\partial}{\partial r}\left(\frac{\delta T}{\Tbar}
\right)\right]-\frac{1}{\sin\theta}
\frac{\partial}{\partial\theta}\left[\sin\theta\frac{\partial}
{\partial\theta}\left(\frac{\delta T}{\Tbar}\right)\right]
=\frac{2f(\theta)}{\Vbar^0}=\sum_{\ell\neq 0}\ellp\tl\pell(\theta),
\eeq
which has the general solutions
\beq
{\delta T\over\Tbar}=\sum_{\ell\neq 0}
[\tl+c_\ell\cos(\kl/\gbar(r))
+s_\ell\sin(\kl/\gbar(r))]\pell(\theta),
\label{eq:tsoln}
\eeq
where $\kl\equiv\sqrt{\ellp/2}$, and $c_\ell$ and $s_\ell$ are 
constants to be determined from boundary conditions.
Notice that since $\gbar(r)\propto r$, the temperature fluctuations
become independent of $r$ asymptotically:
\beq
\lim_{r\to\infty}{\delta T\over\Tbar}=\sum_{\ell\neq 0}
(\tl+c_\ell)\pell(\theta).
\label{eq:tlim}
\eeq
From equations (\ref{eq:tsoln}) and
(\ref{eq:omtr2}) we find
\beq
\delta V_\theta={r^2\Tbar^2\over\Vbar^0}
\sum_{\ell\neq 0}\left[{\tl\over r}+{c_\ell\gbar\over\kl r}
\sin(\kl/\gbar(r))-{s_\ell\gbar\over\kl r}\cos(\kl/\gbar(r))\right]
{d\pell(\theta)\over d\theta}+{A_\theta\over\sin\theta};
\label{eq:Vt}
\eeq
the additional solution proportional to $1/\sin\theta$ is a 
(singular) potential flow with no associated temperature perturbation.
Since $\delta V_\theta=r^2\Tbar\delta U^\theta$, and $\Tbar r=$ constant,
the non-radial velocity component $\delta v^\theta=r\delta U^\theta/\gamma
\propto r^{-1}$ at large
radii (ignoring the singular term in eq.[\ref{eq:Vt}]).
From equation (\ref{eq:norm}) we find that 
\beq
\delta V_r=-{\Vbar^0\over 2}\sum_{\ell\neq 0}
\left\{\tl\ellp
+{2\Tbar^2\over(\Vbar^0)^2}
[c_\ell\cos(\kl/\gbar(r))
+s_\ell\sin(\kl/\gbar(r))]\right\}\pell(\theta);
\label{eq:Vr}
\eeq
here, we ignored a correction term $\propto \Tbar^2/(\Vbar^0)^2$
compared with $\ellp$ in the term $\propto\tl$. To leading order
in $1/\gbar^2$, $\delta V^r=\delta V^0$.
We note that in this limit $\omega_{0\theta}=\omega_{\theta r}=
df(\theta)/d\theta$. In the solutions given above, only the 
terms proportional to $\tl$ represent flows with nonzero vorticity;
the remaining terms are the general solution for perturbed potential
flow.

Although memory of the temperature fluctuations at the inner
boundary is maintained far out in the flow, it is the value of
$V^0=\gamma T$ that determines whether there are any observable
consequences. As a result, if $f(\theta)=0$, so that the flow is 
derivable from a potential, the radiation spectrum detected by
a distant observer is the same as for the background spherical
flow. If $f(\theta)\neq 0$, so that the flow has nonzero
vorticity, distant observers detect a superposition of 
quasi-thermal spectra of the kind derived in Section \ref{obs_spec}. 
While these
results have been derived for small axisymmetric perturbations
about spherically symmetric flow, it is conceivable that when
the perturbations become significant, the observed spectrum
can appear substantially non-thermal.

\section{Baryon Contamination and $\gamma_\infty$}
\label{baryon}

How is the $\mdot \to 0$ limit achieved for nonzero (baryon) 
mass ejection rate $\mdot$? This question raises three 
subsidiary ones: 
(1) For a given $\edot$, what is the critical value of $\edot/\mdot$
above which the flow is essentially the same as for $\mdot=0$?
(2) Since we know that the asymptotic Lorentz factor $\gamma_\infty$ 
is finite for $\mdot=0$ (equation [\ref{am_ginfty}]), and that 
$\gamma_\infty\sim\edot/\mdot$ for relatively large $\mdot$
(small $\edot/\mdot$), does $\gamma_\infty$ grow monotonically 
with increasing $\edot/\mdot$?
(3) What is the largest possible value of $\gamma_\infty$?

Below, we demonstrate that $\gamma_\infty\sim\edot/\mdot$ 
until it reaches a critical value 
\beq
\left(\frac{\edot}{\mdot}\right)_{\rmn{c,M}} \sim
350 (Z/A)^{1/4}(r_i/10^6\mbox{cm})^{1/4}\gamma_{i}^{3/4}(T_i/m_e),
\eeq
at which the radius $r_M\sim(\edot/\mdot)r_i$ where the flow 
becomes matter dominated first moves out to the photosphere.
As $\mdot$ decreases further, we find that $\gamma_\infty$ 
remains virtually constant at $\sim(\edot/\mdot)_{\rmn{c,M}}$ until 
a second critical value
\beq
\left( \frac{\edot}{\mdot} \right)_{\rmn{c,P}} \sim
7\times10^4(Z/A)(r_i/10^6\mbox{cm})(T_i/m_e),
\eeq
above which the densities of electrons and positrons are 
nearly equal once pairs go out of equilibrium.
Between $(\edot/\mdot)_{\rmn{c,P}}$ and 
\beq
\left( \frac{\edot}{\mdot} \right)_{\rmn{c,0}} \sim
5\times10^7(r_i/10^6\mbox{cm})(T_i/m_e)
\eeq
the inertia in $e^\pm$ pairs beyond the photosphere becomes
progressively more important compared with the inertia 
in baryons, and $\gamma_\infty$ grows from 
$\sim (\edot/\mdot)_{\rmn{c,M}}$ to roughly its maximum value,
the $\mdot\to 0$ limit, which applies at $\edot/\mdot\simgreat
(\edot/\mdot)_{\rmn{c,0}}$. Over the entire domain $\edot/\mdot\simgreat
(\edot/\mdot)_{\rmn{c,M}}$, the value of $\gamma_\infty$ only grows 
by a factor $\sim(Am_p/2Zm_e)^{1/4}\sim10$.

\subsection{Equilibrium flow}

First, we review the results 
for an equilibrium flow with baryons (Paczy\'{n}ski 1986).
From conservation of baryon number and energy, it follows that
\beqa
\ndot & \equiv & 4\pi r^2n\gamma v = \mdot/Am_p = \mbox{constant}; 
 \\
\edot & \equiv & 4\pi r^2(\rho+P)\gamma^2 v = \mbox{constant}. 
\eeqa
As before, we get $\gamma\propto r$ and $T\propto 1/r$ in a radiation 
dominated flow. This behavior is maintained until $\rho+p\sim nAm_p$, 
which occurs at a radius $r_M\approx (\edot/\mdot)(r_i/\gamma_i)$ where
$\gamma\sim\edot/\mdot$.
In an equilibrium flow at sufficiently low $T$
the optical depth is essentially 
$\tau \sim Zn\sigt r/\gamma$, resulting in 
\beq
\tau \approx \frac{Z\ndot^4(Am_p)^3\gamma_i\sigt}{4\pi r_i\edot^3}
\times\left\{ 
\begin{array}{ll}
(r_M/r)^3 & \quad\mbox{for}\quad r_M>r \\
(r_M/r) & \quad\mbox{for}\quad r_M<r.
\end{array}
\right.
\eeq
A critical value of $\edot/\mdot$ is defined as the value which 
gives $\tau=1$ at $r=r_M$, i.e. $r_{\rmn{ph}}=r_M$. With $\rho+P\sim
(11/4)(4/3)(\pi^2/15)T_{i}^{4}$ at $r_i$, this 
gives
\beq
\left( \edot/\mdot \right)_{\rmn{c,M}} \sim
350\left( \frac{Z}{A} \right)^{1/4}\left( \frac{r_i}{10^6\mbox{cm}}
\right)^{1/4}\gamma_{i}^{3/4}\frac{T_i}{m_e}.
\eeq
The optical depth can then be rewritten as
\beq
\tau \approx \left[ \frac{ \left( \edot/\mdot \right)_{\rmn{c,M}}}{\edot/\mdot}
\right]^4
\times\left\{ 
\begin{array}{ll}
(r_M/r)^3 & \quad\mbox{for}\quad r_M>r \\
(r_M/r) & \quad\mbox{for}\quad r_M<r.
\end{array}
\right.
\eeq
If $\edot/\mdot < (\edot/\mdot)_{\rmn{c,M}}$, $r_M$ is inside the photosphere
and we expect there to be little acceleration outside $r_M$, so
the asymptotic Lorentz factor is $\edot/\mdot$.
If, on the other hand, $\edot/\mdot > (\edot/\mdot)_{\rmn{c,M}}$, then $r_M$ 
is outside the photospheric radius 
$r_{\rmn{ph}} \approx r_M\left[ (\edot/\mdot)_{\rmn{c,M}}/(\edot/\mdot) \right]^{4/3}$,
and the Lorentz factor at $r_{\rmn{ph}}$ is
$\gamma_{\rmn{ph}}\approx \gamma_M(r_{\rmn{ph}}/r_M) \approx 
(\edot/\mdot)^{4/3}_{\rmn{c,M}}(\edot/\mdot)^{-1/3}$.
The acceleration of the flow by the radiation flux beyond $r_{\rmn{ph}}$ is 
found in a similar way as for the non-equilibrium pair wind, using
\beq
Am_p\frac{d\gamma}{dr} \approx Z\sigt\fo(r) = 
\frac{4\pi}{3}\left( \frac{r_{\rmn{ph}}\gamma_{\rmn{ph}}}{r\gamma} \right)^2 \sigt
ZI_{\rmn{ph}}\left[ 1-\left(\frac{r_{\rmn{ph}}\gamma}{r\gamma_{\rmn{ph}}}\right)^4 \right].
\eeq
For $r_M>r_{\rmn{ph}}$ this results in a final Lorentz factor 
$\gamma_\infty \sim (\edot/\mdot)_{\rmn{c,M}} < \edot/\mdot$.

The use of a free streaming form for the radiation field for
$\tau<1$ implies that the radiative component of $\edot$ 
is constant in this regime. This is a good approximation when
$\edot/\mdot \simgreat (\edot/\mdot)_{\rmn{c,0}}$ and the energy 
content of the baryons is negligible. When
$\edot/\mdot \simless (\edot/\mdot)_{\rmn{c,M}}$, the radiation 
energy is much smaller than the baryon energy outside the photosphere,
and so the 
details of the radiation field are not dynamically important.
In the intermediate regime, radiative acceleration increases 
the baryonic energy, causing a corresponding decrease in the
radiative energy unaccounted for when using the free streaming
approximation. 
The error thus made can be estimated as
\beq
\Delta\dot{E}_{\rmn{baryon}}/\dot{E} = (\gamma_\infty-\gamma_{\rmn{ph}})\mdot/\edot
\sim \frac{(\edot/\mdot)_{\rmn{c,M}}}{\edot/\mdot}\left\{ 
1-\left[ \frac{(\edot/\mdot)_{\rmn{c,M}}}{\edot/\mdot} \right]^{1/3} \right\}.
\label{barerr}
\eeq
The maximum $\Delta\dot{E}_{\rmn{baryon}}/\dot{E}$ is $\sim 0.1$ and 
occurs at $\edot/\mdot = (4/3)^3(\edot/\mdot)_{\rmn{c,M}}$.
Numerically we find the maximum gain in baryonic energy to 
be about 20 per cent of the total flow energy, with a functional 
dependence on $\edot/\mdot$ following roughly the estimate 
in equation (\ref{barerr}). Since the error made in assuming 
a free streaming radiation field is relatively small as far as
the dynamics is concerned and only affects a small interval in
$\edot/\mdot$, we adopted this simplification in our
numerical calculations. The resulting values of $\gamma_\infty$ 
are very accurate for $\edot/\mdot$ well above and well below 
$(\edot/\mdot)_{\rmn{c,M}}$, and should be correct to $\approx 10-20$ per cent
for $\edot/\mdot\sim(\edot/\mdot)_{\rmn{c,M}}$.

\subsection{Baryons and non-equilibrium effects}

With baryons in the flow, charge neutrality 
requires that the electron number density $n_-$, the positron
number density $n_+$, and the baryon number density $n$ obey 
$n_-=n_+ + Zn$. The positron number density $n_+$ is then 
found from
\beq
\frac{1}{r^2}\frac{d}{dr}\left( r^2n_+\gamma v \right) =
-\langle\sigma_{\rmn{ann}}v\rangle \left[ \left( n_++Zn\right)n_+ - n_{\rmn{eq}}^{2} \right],
\eeq
where, as before,
$n_{\rmn{eq}}^{2}=(n_+n_-)_{\rmn{eq}}$.
At large $\mdot$, the excess of electrons over positrons is always large
at $T<m_e$. The positron density first becomes important at $r_{\rmn{eq}}$ when
\beq
\frac{\edot}{\mdot}\sim
\frac{Zm_e}{Am_p}\frac{\gamma_iT_i}{m_e}\frac{C}{(\ln C)^4}\approx
3\times10^3\frac{Z}{A}\frac{T_i}{m_e}\left( \frac{r_i}{10^6\mbox{cm}} \right)
\eeq
where $C\equiv\pi(e^4/m_{e}^{2})m_{e}^{3}r_i/\gamma_i\approx
4.4\times10^{12}(r_i/10^6\mbox{cm})\gamma_{i}^{-1}$; positrons first 
become important at the photosphere and beyond when
\beq
\frac{\edot}{\mdot}\approx\left( \frac{\edot}{\mdot} \right)_{\rmn{c,P}}\approx
\frac{Zm_e}{Am_p}\frac{\gamma_iT_i}{m_e}\frac{C}{(\ln C)^3}.
\eeq
For $\edot/\mdot>(\edot/\mdot)_{\rmn{c,P}}$, the inertial mass density 
outside the photosphere is 
\beq
\rho\approx Am_pn+2m_en_+\equiv n_+m_{\rmn{eff}}
\eeq
where the effective mass is approximately 
\beq
m_{\rmn{eff}}\approx 2m_e\left[ 
1+(\mdot/\edot)(T_i/m_e)C\gamma_i/4(\ln C)^3 \right],
\eeq
which varies between $\approx Am_p/Z$ for $\edot/\mdot\sim(\edot/\mdot)_{\rmn{c,P}}$
and $2m_e$ for 
\beq
\frac{\edot}{\mdot}>\left( \frac{\edot}{\mdot}\right)_{\rmn{c,0}} \approx
\frac{T_i}{4m_e}\frac{C\gamma_i}{(\ln C)^3}.
\eeq
Notice that $(\edot/\mdot)_{\rmn{c,0}}\sim(Am_p/2Zm_e)(\edot/\mdot)_{\rmn{c,P}}$. 

The acceleration caused by the radiation force for $\tau<1$ can 
be estimated using
\beq
m_{\rmn{eff}}\frac{d\gamma}{dr} =
\frac{4\pi}{3}\left( \frac{r_{\rmn{ph}}\gamma_{\rmn{ph}}}{r\gamma} \right)^2 \sigt
I_{\rmn{ph}}\left[ 1-\left(\frac{r_{\rmn{ph}}\gamma}{r\gamma_{\rmn{ph}}}\right)^4 \right],
\eeq
resulting in
\beq
\gamma_\infty \sim \frac{ (Am_p/2Zm_e)^{1/4}(\edot/\mdot)_{\rmn{c,M}} }
{ \left[ 1 + (\mdot/\edot)(T_i/m_e)C\gamma_i/4(\ln C)^3 \right]^{1/4} }.
\label{gam_bar}
\eeq
Equation (\ref{gam_bar}) shows that $\gamma_\infty$ rises slowly 
($\sim[\edot/\mdot]^{1/4}$) from $\sim(\edot/\mdot)_{\rmn{c,M}}$ at 
$\edot/\mdot\sim(\edot/\mdot)_{\rmn{c,P}},$ to
$\sim(Am_p/2Zm_e)^{1/4}(\edot/\mdot)_{\rmn{c,M}}$ at 
$\edot/\mdot\gg(\edot/\mdot)_{\rmn{c,0}}$. Since 
$\gamma_\infty\sim(\edot/\mdot)_{\rmn{c,M}}$ for 
$(\edot/\mdot)_{\rmn{c,P}}\simgreat\edot/\mdot\simgreat(\edot/\mdot)_{\rmn{c,M}}$,
the total increase in $\gamma_\infty$ for {\em all} $\edot/\mdot\simgreat
(\edot/\mdot)_{\rmn{c,M}}$ is a factor $\sim(Am_p/2Zm_e)^{1/4}\sim6$. This 
alleviates any `fine-tuning' problem necessary to produce large
asymptotic Lorentz factors (although proportionality to $T_i/m_e$ remains):
Values within an order of magnitude of one another are found as long as
$\edot/\mdot$ is sufficiently large.

\begin{figure}
\begin{picture}(200,225)(-125,0)
\put(0,3){\epsfxsize=3.0in\epsffile{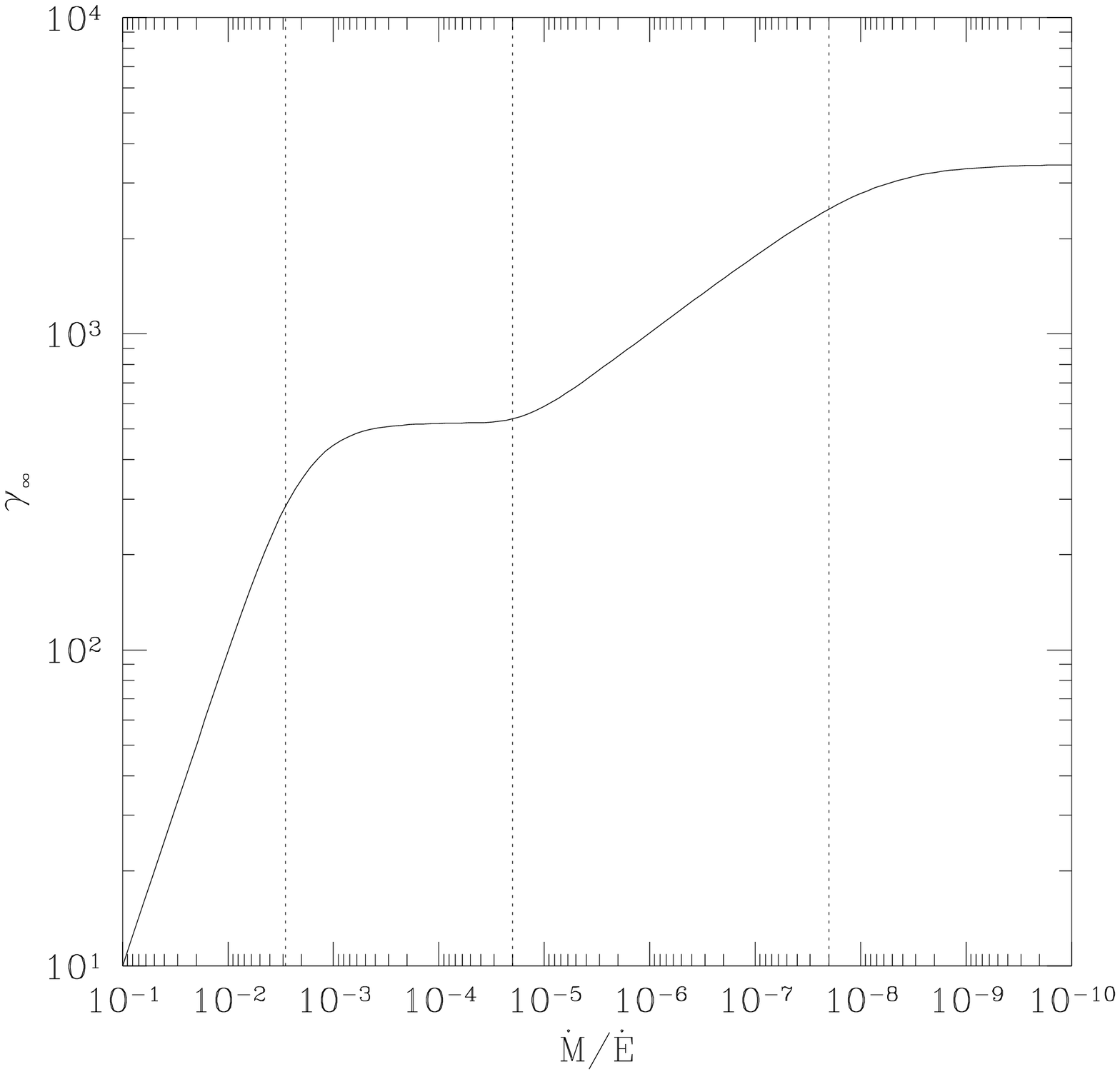}}
\end{picture}
\caption{Asymptotic Lorentz factor as a function
of $\mdot/\edot$, the ratio of the mass injection rate to the 
energy injection rate. The dotted vertical lines correspond to 
the transition points $(\edot/\mdot)_{\rmn{c,M}}, (\edot/\mdot)_{\rmn{c,P}}$,
and $(\edot/\mdot)_{\rmn{c,0}}$. 
In this example, $T_i/m_e=1, \gamma_i=2,
\siga/\sigt=10^{-3}, \sigs/\sigt=1$ and $A=Z=1$. 
\label{moe_plot}}
\end{figure}

This qualitative behavior is seen in the numerical solution:
$\gamma_\infty$ is shown as a function of $\mdot/\edot$ in 
Fig. \ref{moe_plot}, clearly showing the four distinct 
regions discussed above. 
The transitions between the regions are
seen to correspond to the critical points $(\edot/\mdot)_{\rmn{c,M}}$,
$(\edot/\mdot)_{\rmn{c,P}}$, and  
$(\edot/\mdot)_{\rmn{c,0}}$, shown as dotted vertical lines in
Fig. \ref{moe_plot}.

\section{Discussion}
\label{discussion}

In order to make our treatment of the relativistic $e^\pm\gamma$
wind somewhat more complete, we address briefly
(and largely qualitatively) two additional questions:
(1) What happens if the initial temperature is much higher than
   the electron mass, e.g. comparable to the mass of muons or
   even nucleons?
(2) Do weak magnetic fields alter the flow significantly?

\subsection{The effects of $T_i\gg m_e$}

When $T\sim m_\mu$, the muons are in equilibrium with the 
radiation, $\mu^+ + \mu^- \leftrightarrow \gamma + \gamma$.
As $T$ drops below a few percent of $m_\mu$, the muon pair 
annihilations freeze out in the same way as the electron--positron
pair annihilations 
do at $T\sim 0.05m_e$. Going through the same calculations as 
for the electrons, we find that
$r_{\rmn{eq,e}}/r_{\rmn{eq},\mu} \sim {m_\mu}/{m_e}$,
and that $\rho_\mu\approx \rho_e$ for $r>r_{\rmn{eq,e}}$ assuming that
the muons have not decayed before reaching $r_{\rmn{eq},\mu}$. 
(For $r_{\rmn{eq},\mu}\ll r\ll r_{\rmn{eq,e}}$,
$m_e\ll T \ll m_\mu$ and therefore $\rho_\mu\ll \rho_e$ in this regime.)
However, the mean lifetime of a muon is only $t_\mu\sim2.2\times 10^{-6}$
seconds. They will therefore be significantly abundant
at $r_{\rmn{eq,e}}$ only if 
$t_\mu > \int_{r_{\rmn{eq},\mu}}^{r_{\rmn{eq,e}}} dr/\gamma(r)$, i.e., if
$\gamma_i > 80 (r_i/10^6\mbox{cm})$.
For $\gamma_i$ smaller than this, the presence of muons in the 
flow will not have any significant influence on its dynamics.
On the other hand, if $\gamma_i$ is large enough for the muons 
to survive until $\gamma$ reaches $\gamma_{\infty}$, then they will reduce 
the asymptotic Lorentz factor slightly because of their added 
contribution to the inertia of the flow. (Demanding $t_\mu > \int_{r_{\rmn{eq},\mu}}^{r_{\gamma,e}}dr/\gamma(r)$ requires $\gamma_i>1.9\times10^2[r_i/10^6\mbox{cm}]$.)
For $T\ll m_e$, equation (\ref{eeq21_5}) reduces to 
\beq
(\rho_e+\rho_\mu)\frac{d\ln\gamma}{d\ln r} \approx
-\frac{r}{\gamma}G_{0}^{r} \approx \frac{r}{\gamma}2n_e(\sigma_a + \sigma_s)F_0
\eeq
when muons are abundant.
(Since $\sigma_T\propto m^{-2}$, we may neglect the muon contribution 
to the scattering cross section.)
Thus the radiation force is approximately equal 
to that of a flow with only electrons
and positrons,
whereas the inertia is doubled since 
$\rho_e+\rho_\mu\approx 2\rho_e \approx 2m_en_e$ 
in the region outside the photosphere.
In the notation of Section \ref{am_tault1}, then, 
$\Lambda\to\Lambda/2$. And since $\gamma_\infty\sim\gamma_{\rmn{ph}}\Lambda^{1/4}$,
the muons will reduce the asymptotic Lorentz factor by a factor $2^{1/4}\approx
1.2$, provided that most of them have not decayed by the end of the 
acceleration epoch.
Ultimately, the relativistic muons decay, resulting in an additional 
non-thermal population of electrons.

In the arguments above we implicitly assumed that muons and electrons 
are coupled. Since their coupling via photons is very weak, the relevant
remaining mechanism is through Coulomb scattering. The time-scale 
for Coulomb interaction is roughly (See e.g. Spitzer 1978)
\beq
t_c \sim \left\{ n_\mu\sqrt{\frac{T}{m_e}}\frac{e^4}{T^2} 
{4\ln \left[ (3/2e^3)\sqrt{T^3/\pi n_e} \right]} \right\}^{-1}.
\label{dceq3}
\eeq
Comparing this to the expansion time-scale, $t_{\rmn{exp}}\sim r/\gamma$,
we get
${t_c}/{t_{\rmn{exp}}} \sim 7.2\times10^{-5}(m_e/T)^{3/2}$, which is
less than unity for $T/m_e>1.7\times10^{-3}$.
Recall that $T_{\rmn{ph}}/m_e \sim 0.037$ and $T_\gamma/m_e \sim 8\times 10^{-4}$;
this indicates that the electrons and muons are 
marginally coupled for most of the acceleration regime $r_{\rmn{ph}} < r < r_\gamma$.
The reduction of $\gamma_\infty$ by a factor 1.2 therefore represents
an estimate for the maximum change in the asymptotic Lorentz factor
caused by the presence of muons in a flow with very high $\gamma_i$
(or small $r_i$).

For high temperatures the inner portion of the wind can be optically 
thick to neutrinos as well. The cross section for electron--neutrino
scattering is $\sigma_\nu \sim \sigma_0(T/m_e)^2$, where $\sigma_0 = 
1.76\times10^{-44}\mbox{cm}^2$. The corresponding optical depth is 
\beq
\tau_\nu(r) \sim \int_{r}^{\infty}drn_e(r)\sigma_\nu(r)/\gamma(r) 
\sim 1.4\times 10^{-8}\frac{1}{\gamma_i}\frac{r_i}{10^6\mbox{cm}}
\left(\frac{T}{m_e}\right)^5;
\eeq
Consequently, $\tau_\nu>1$ for $T/m_e > 37\gamma_{i}^{1/5}(r_i/10^6\mbox{cm})^{-1/5}$.
Neutrinos are kept in thermal equilibrium via reactions like 
$\nu_e + \bar{\nu}_e \leftrightarrow e^- + e^+$, whose rate exceeds 
the expansion rate for $T/m_e > 31\gamma_{i}^{1/5}(r_i/10^6\mbox{cm})^{-1/5}$.

If $T$ increases further we have to include not only muons in the flow, 
but also mesons. However, since mesons are very short--lived, their 
dynamical effect on the flow will be negligible. But for initial 
temperatures high enough for nucleon--anti-nucleon pairs to exist, there 
is the possibility that the surviving fraction (after annihilations
freeze out) will result in a significant baryon loading in the wind.
The cross section for nucleon--anti-nucleon annihilations is 
$\langle \sigma v\rangle \sim 1/m_{\pi}^{2}$, resulting in a 
freeze--out temperature of $T_{\rmn{eq,nucl}}/m_p \sim 0.026$.
This results in negligible baryon loading in the wind:
$\rho_{\rmn{nucl}}/\rho_{e^\pm}\sim 4\times10^{-6}$ for $r\gg r_{\rmn{eq,e}}$.

\subsection{Effects of weak magnetic fields}

In this subsection we will estimate how weak magnetic fields
can affect the flow.
The `background' wind is assumed to expand radially, and 
we consider the effects of radial and tangential fields.

In Section \ref{pert} we presented a general treatment of perturbations
of a flow with equation of state that depends only on temperature.
When magnetic fields are present, and the electric field vanishes in
the rest frame of the flow, the equation of entropy conservation,
$(\sigma U^\mu)_{;\mu}=0$ remains true, but the remaining fluid 
equations are modified to
\beq
V^\nu\omega_{\mu\nu}={F_{\mu\lambda}J^\lambda\over Q(T)}
\label{eq:bforce}
\eeq
where $F_{\mu\lambda}$ is the Maxwell tensor, and $J^\lambda$ 
is the current density four vector. 

We can use equation (\ref{eq:bforce}) to examine the effects on the
flow of a magnetic field; to do so requires solving Maxwell's
equations simultaneously. A particularly simple example is a
radially directed magnetic field, as might arise, for example, if
the flow originates on a magnetized star, and pulls magnetic
field lines outward along with it. In that case, it is easy to 
show that, assuming axisymmetry, the magnetic field strength
is proportional to $b(\theta)/r^2$, where $b(\theta)$ is arbitrary.
To lowest order, the electric field associated with a radial
magnetic field vanishes in the frame of a stationary observer as
well as in the frame comoving with the flow.

Under such conditions, the magnetic
field engenders {\it no} change in $\gamma T$ to first order. This
follows from the $\mu=r$ component of equation (\ref{eq:bforce}),
which implies $\Vbar^0\omega_{0r}=0$, or $\partial\delta V_0/\partial
r=0$. (A similar conclusion follows from the $\mu=0$ component of
eq. [\ref{eq:bforce}].)
Thus, the only perturbations to $\delta V_0$ are those imposed
at the inner radius of the flow, as discussed in Section \ref{pert}; 
none are created by the field.
Radial magnetic fields drive non-radial flows whose amplitudes
decline with radius even outside the photosphere (Grimsrud 1998).

Changes in $\gamma T$ require a non-radial magnetic field. The condition
of perfect conductivity, $\Evec+\vvec\times\Bvec=0$, and Faraday's law
imply $\Evec=-\gradvec\psi=-\vvec\times\Bvec$ in steady state; for 
axisymmetry we find, to lowest order,
\beq
\Bvec=-{\ephi\over rv(r)}\dpsi\qquad\qquad
\Evec=-{\eth\over r}\dpsi.
\label{eq:ebtang}
\eeq
Associated with ordered tangential fields is a Poynting flux in the
stationary frame,
\beq
\Svec={\er\over 4\pi r^2v(r)}\left(\dpsi\right)^2;
\eeq
there is no Poynting flux for radially directed $\Bvec$ to lowest
order (since $\Evec=0$), and there is, of course, no Poynting flux
in the comoving frame. Note that $4\pi r^2\Svec$ decreases with
increasing radius as $v(r)\to 1$ from below. For these fields 
the $\mu=0$ component of equation (\ref{eq:bforce}) implies
\beq
{\partial\delta V_0\over\partial r}
={1\over 4\pi r^2Q(\Tbar)\Vbar^r}\left(\dpsi\right)^2
{d\over dr}\left({1\over v}\right)
=-{1\over 4\pi r^2Q(\Tbar)\Vbar^rv^3\gbar^3}\left(\dpsi\right)^2
{d\gbar\over dr};
\label{eq:bf0}
\eeq
the $\mu=r$ component of equation (\ref{eq:bforce}) yields an
identical result. In the extreme relativistic limit, where
$\gbar\propto r$ and $Q(\Tbar)\propto \Tbar^2\propto r^{-2}$, 
equation (\ref{eq:bf0})
has the solution
\beq
\delta V_0={1\over 8\pi r^2Q(\Tbar)\gbar^2\Vbar^0}\left(\dpsi
\right)^2=
{\vert\Bvec\vert^2\over 8\pi Q(\Tbar)\gbar^2\Vbar^0},
\eeq
so $\delta V_0\propto r^{-2}$. Although tangential fields alter
$\gamma T$, the perturbation peaks near the lift-off radius, and
decreases far out in the flow. We therefore expect little or no
effect on the observed radiation spectrum as a consequence of 
such fields. There is additional radial acceleration of the
$e^\pm$ pairs outside the photosphere as a consequence of the
tangential field; we estimate the change in bulk Lorentz
factor of the pairs to be $\Delta\gamma\approx (d\psi(\theta)/
d\theta)^2/4m_e\dot N\gamma_{ph}^2$ if pair annihilation can be
neglected outside the photosphere, so the pair loss rate
$\dot N$ becomes independent of radius.

\section{Conclusions}
\label{conclusions}

Radiation energy can escape from a fireball in two different 
ways: If there is significant baryon contamination present, 
much of the energy will be converted into bulk kinetic energy
(Shemi \& Piran 1990).
However, when the expanding atmosphere has swept up a significant
amount of surrounding matter, kinetic energy can be converted 
into escaping radiation at the resulting shock front
(M\'{e}sz\'{a}ros \& Rees 1993).
In a similar way, internal shocks due to a non-uniform 
velocity can convert kinetic energy into radiation
(Rees \& M\'{e}sz\'{a}ros 1994).
The other mechanism for radiation escape is more direct:
If the particle content is small, the fireball can become 
optically thin before being matter dominated.

In this paper we have considered an extreme case of the latter
possibility, in which there are no baryons present.
The opacity is then due to electron--positron pairs, resulting 
in a very large optical depth for temperatures 
greater than the electron mass. Further out in the flow
the temperature decreases, pair creation is suppressed and 
annihilations freeze in; this results in a small but
non-negligible amount of surviving pairs.
The radiation force acting on the particles accelerates
the pairs considerably, even after the flow has become 
optically thin.
We found that the Lorentz factor of the flow approaches the constant
value $\gamma_{\infty} \sim 1.4\times 10^{3} \gamma_{i}^{3/4} 
(T_i/m_e)[(\siga/\sigt)+(\sigs/\sigt)]^{1/4}$ only when the optical depth
falls below $\tau_{\gamma} \sim 1.7 \times 10^{-5} \gamma_{i}^{3/4}$.
This increases the asymptotic energy content of the pairs by a 
large factor, their fraction of the total energy approaching 
$\sim 8.5 \times 10^{-6} \gamma_{i}^{3/4} [(\siga/\sigt)+(\sigs/\sigt)]^{1/4}$.
The flow is always radiation dominated for reasonable values of 
the input parameters.

Even if the initial temperature were much higher than the 
electron mass, the resulting flow would not deviate
significantly from an $e^\pm\gamma$ wind. If there are 
muons present, they will decay before the electron--positron
annihilations freeze out, unless $\gamma_i$ is very large.
And even for $\gamma_i$ high enough for the muons to survive 
until far outside the photosphere, their added inertia will only
reduce the asymptotic Lorentz factor by at most 20 per cent.
For even higher $T_i$, one may have nucleon--anti-nucleon pairs
present in the flow. However, nucleon--anti-nucleon annihilations 
freeze out at a relatively low temperature, thus causing the
baryon contamination in the flow to be negligible.

The photon distribution function in the comoving frame is very close
to that of blackbody radiation.
This is because $\gamma \propto r$ and $\gamma T \approx$ constant
are excellent approximations in the flow until 
$r=r_{\gamma}$ where the optical depth is $\tau_{\gamma} \sim
1.7\times 10^{-5} \gamma_{i}^{3/4}$. Practically all the observed radiation
therefore originates from a region where these two 
approximations hold.
As was discussed in Section \ref{spec}, the conditions
$\gamma \propto r$ and $\gamma T =$ constant imply that the
equation of radiative transfer is solved by a blackbody 
distribution function in the comoving frame of the flow. 
The spectrum seen by an observer in the lab frame will 
deviate somewhat from a blackbody in that it has a 
broader peak and a shallower slope at low photon energies.
Such spectra are {\em not} typical of observed $\gamma-$ray 
bursts, which are characterized by flat fluxes for 
logarithmic energy intervals
(e.g. Schaefer et al. 1992, 1994; Kouveliotou 1994).
A superposition of 
quasi-thermal spectra from numerous source regions radiating
independently (but with different physical parameters)
might produce flat spectra.
The non-radial perturbation calculations discussed in 
Section \ref{pert} lend partial support to this idea.

Ordered magnetic fields (whose energy content is small compared to that of 
the flow) will 
make the temperature and the velocity of the flow anisotropic
and may enhance the bulk Lorentz factor of $e^\pm$ pairs
beyond the photosphere, 
but do not
affect the spectrum seen by a distant observer significantly.

The results obtained for zero baryon number also apply when the
baryon loading is sufficiently small. For very large baryon loading,
the flow becomes matter dominated at optical depths larger than one,
and in this case the asymptotic Lorentz factor 
$\gamma_\infty\sim\edot/\mdot$.
As $\edot/\mdot$ increases above $(\edot/\mdot)_{\rmn{c,M}}\sim
350(Z/A)^{1/4}(r_i/10^6\mbox{cm})^{1/4}\gamma_{i}^{3/4}T_i/m_e$,
the asymptotic Lorentz factor at first levels off at 
$\gamma_\infty\sim(\edot/\mdot)_{\rmn{c,M}}$. For still larger
$\edot/\mdot\simgreat(\edot/\mdot)_{\rmn{c,P}}\sim7\times10^4(Z/A)
(r_i/10^6\mbox{cm})(T_i/m_e)$, the asymptotic Lorentz factor 
rises $\sim(\edot/\mdot)^{1/4}$, until $\edot/\mdot\sim(\edot/\mdot)_{\rmn{c,0}}
\sim(Am_p/2Zm_e)(\edot/\mdot)_{\rmn{c,P}}$. The $\mdot\to 0$ limit 
applies for $\edot/\mdot>(\edot/\mdot)_{\rmn{c,0}}\sim
5\times10^7(r_i/10^6\mbox{cm})(T_i/m_e)$; in this regime,
$\gamma_\infty\sim(Am_p/2Zm_e)^{1/4}(\edot/\mdot)_{\rmn{c,M}}$.
The fraction of the total wind luminosity that emerges in the 
form of bulk kinetic energy falls below one at $\edot/\mdot\sim
(\edot/\mdot)_{\rmn{c,M}}$, and decreases monotonically until 
asymptoting to a finite value $\sim10^{-5}\gamma_{i}^{3/4}$ as
$\mdot\to0$ (see equation [\ref{am_eratio}]).
Thus, for all $\edot/\mdot>(\edot/\mdot)_{\rmn{c,M}}$, the asymptotic
Lorentz factor varies by a factor of only $\sim(Am_p/2Zm_e)^{1/4}\sim6$.
The maximum possible $\gamma_\infty$ is the value found for 
$\mdot=0$ and is finite.
Although we have only considered steady winds here, it seems likely
that similar results would hold for fireballs originated impulsively.

\acknowledgments

This research was supported in part by NSF grant AST 93-15375 and 
NASA grant NAG 5-3097. OMG thanks Nansenfondet for financial support.


\begin{references}
\reference{abramowicz} Abramowicz M. A., Novikov I. D., Paczy\'{n}ski B., 1991, ApJ, 369, 175
\reference{fenimore} Fenimore E. E., 1997, in 
Olinto A., Friemann J., Schramm D., eds,
Proc. 18th Texas Symposium on Relativistic Astrophysics
\reference{fuller} Fuller G. M., Shi X., 1997, astro-ph/9711020
\reference{goodman} Goodman J., Dar A., Nussinov S., 1987, ApJ, 314, L7
\reference{ole} Grimsrud O. M., 1998, PhD thesis, Cornell University
\reference{hummer} Hummer D. G., Rybicki G. B., 1971, MNRAS, 152, 1
\reference{kobay} Kobayashi S., Piran T., Sari R., 1997, ApJ, 490, 92
\reference{kouv} Kouveliotou C., 1994, ApJS, 92, 637
\reference{lw} Lee B. W., Weinberg S., 1977, Phys. Rev. Lett., 39, 165
\reference{rees_shock} M\'{e}sz\'{a}ros P., Rees M. J., 1993, ApJ, 405, 278
\reference{mandm} Mihalas D., Mihalas B. W., 1984, Foundations of Radiation Hydrodynamics. Oxford Univ. Press, Oxford
\reference{naryan} Naryan R., Paczy{\'n}ski B., Piran T., 1992, ApJ, 395, L83
\reference{bp86} Paczy\'{n}ski B., 1986, ApJ, 308, L43
\reference{bp90} Paczy\'{n}ski B., 1990, ApJ, 363, 218
\reference{bp97} Paczy\'{n}ski B., 1997, astro-ph/9712123
\reference{turok} Pen U., Loeb A., Turok N., 1997, astro-ph/9712178
\reference{piran97} Piran T., 1997, in 
Bahcall J., Ostriker J.P., eds,
Unsolved Problems in Astrophysics. Princeton Univ. Press, Princeton
\reference{reesshock} Rees M. J., M\'{e}sz\'{a}ros P., 1992, MNRAS, 258, L41
\reference{rees} Rees M. J., M\'{e}sz\'{a}ros P., 1994, ApJ, 430, L93
\reference{randl} Rybicki G. B., Lightman A. P., 1979, Radiative Processes in Astrophysics. John Wiley \& Sons, New York
\reference{scaefer92} Schaefer B. E. et al., 1992, ApJ, 393, L51
\reference{scaefer94} Schaefer B. E. et al., 1994, ApJS, 92, 285
\reference{shemi_matter} Shemi A., Piran T., 1990, ApJ, 365, L55
\reference{spitzer} Spitzer L., 1978, Physical Processes in the 
Interstellar Medium. John Wiley \& Sons, New York
\reference{svensson} Svensson R., 1982, ApJ, 258, 321
\reference{waxman} Waxman E., 1997, ApJ, 485, L5
\reference{waxetal} Waxman E., Kulkarni S. R., Frail D. A., 1998, ApJ, 497, 288
\reference{wijersetal} Wijers R., Rees M. J., M\'{e}sz\'{a}ros P., 1997, MNRAS 288, L51 
\reference{weinberg} Weinberg S., 1972, Gravitation and Cosmology. John Wiley \& Sons, New York
\end{references}
\end{document}